




\font\twelverm=cmr10 scaled 1200   \font\twelvei=cmmi10 scaled 1200
\font\twelvesy=cmsy10 scaled 1200  \font\twelveex=cmex10 scaled 1200
\font\twelvebf=cmbx10 scaled 1200  \font\twelvesl=cmsl10 scaled 1200
\font\twelvett=cmtt10 scaled 1200  \font\twelveit=cmti10 scaled 1200
\font\twelvesc=cmcsc10 scaled 1200
\skewchar\twelvei='177   \skewchar\twelvesy='60


\def\twelvepoint{\normalbaselineskip=12.4pt plus 0.1pt minus 0.1pt
  \abovedisplayskip 12.4pt plus 3pt minus 9pt
  \belowdisplayskip 12.4pt plus 3pt minus 9pt
  \abovedisplayshortskip 0pt plus 3pt
  \belowdisplayshortskip 7.2pt plus 3pt minus 4pt
  \smallskipamount=3.6pt plus1.2pt minus1.2pt
  \medskipamount=7.2pt plus2.4pt minus2.4pt
  \bigskipamount=14.4pt plus4.8pt minus4.8pt
  \def\rm{\fam0\twelverm}          \def\it{\fam\itfam\twelveit}%
  \def\sl{\fam\slfam\twelvesl}     \def\bf{\fam\bffam\twelvebf}%
  \def\mit{\fam 1}                 \def\cal{\fam 2}%
  \def\sc{\twelvesc}               \def\tt{\twelvett}
  \def\sf{\twelvesf}
  \textfont0=\twelverm \scriptfont0=\tenrm
	\scriptscriptfont0=\sevenrm
  \textfont1=\twelvei  \scriptfont1=\teni
	\scriptscriptfont1=\seveni
  \textfont2=\twelvesy \scriptfont2=\tensy
	\scriptscriptfont2=\sevensy
  \textfont3=\twelveex \scriptfont3=\twelveex
	\scriptscriptfont3=\twelveex
  \textfont\itfam=\twelveit
  \textfont\slfam=\twelvesl
  \textfont\bffam=\twelvebf \scriptfont\bffam=\tenbf
  \scriptscriptfont\bffam=\sevenbf
  \normalbaselines\rm}


\def\beginlinemode{\endmode
  \begingroup\parskip=0pt \obeylines\def\\{\par}
	\def\endmode{\par\endgroup}}

\def\beginparmode{\endmode
  \begingroup \def\endmode{\par\endgroup}}
\let\endmode=\par
{\obeylines\gdef\
{}}

\newcount\firstpageno
\firstpageno=2
\footline={\ifnum\pageno<\firstpageno{\hfil}\else{\hfil\twelverm
	\folio\hfil}\fi}

\def\raggedcenter{\leftskip=4em plus 12em \rightskip=\leftskip
  \parindent=0pt \parfillskip=0pt \spaceskip=.3333em
  \xspaceskip=.5em \pretolerance=9999 \tolerance=9999
  \hyphenpenalty=9999 \exhyphenpenalty=9999 }

\hsize=6.5truein
\hoffset=0pt
\vsize=8.9truein
\voffset=0pt
\parskip=\medskipamount

\twelvepoint            

\overfullrule=0pt       

\def\head#1{
  \goodbreak\vskip 0.5truein
  {\immediate\write16{#1}
   \raggedcenter \uppercase{#1}\par}
   \nobreak\vskip 0.25truein\nobreak}

\def\subhead#1{
  \vskip 0.25truein
  {\raggedcenter {#1} \par}
   \nobreak\vskip 0.25truein\nobreak}


\def\refto#1{$^{#1}$}

\def\references
  {\head{References}
   \beginparmode
   \frenchspacing \parindent=0pt \leftskip=1truecm
   \parskip=8pt plus 3pt \everypar{\hangindent=\parindent}}

\gdef\refis#1{\item{#1.\ }}

\def\endreferences{\beginparmode}

\def\ref#1{Ref.~#1}
\def\Ref#1{Ref.~#1}
\def\[#1]{[\cite{#1}]}
\def\cite#1{{#1}}
\def\(#1){(\call{#1})}
\def\call#1{{#1}}


\catcode`@=11
\newcount\tagnumber\tagnumber=0

\immediate\newwrite\eqnfile
\newif\if@qnfile\@qnfilefalse
\def\write@qn#1{}
\def\writenew@qn#1{}
\def\w@rnwrite#1{\write@qn{#1}\message{#1}}
\def\@rrwrite#1{\write@qn{#1}\errmessage{#1}}

\def\taghead#1{\gdef\t@ghead{#1}\global\tagnumber=0}
\def\t@ghead{}

\expandafter\def\csname @qnnum-3\endcsname
  {{\t@ghead\advance\tagnumber by -3\relax\number\tagnumber}}
\expandafter\def\csname @qnnum-2\endcsname
  {{\t@ghead\advance\tagnumber by -2\relax\number\tagnumber}}
\expandafter\def\csname @qnnum-1\endcsname
  {{\t@ghead\advance\tagnumber by -1\relax\number\tagnumber}}
\expandafter\def\csname @qnnum0\endcsname
  {\t@ghead\number\tagnumber}
\expandafter\def\csname @qnnum+1\endcsname
  {{\t@ghead\advance\tagnumber by 1\relax\number\tagnumber}}
\expandafter\def\csname @qnnum+2\endcsname
  {{\t@ghead\advance\tagnumber by 2\relax\number\tagnumber}}
\expandafter\def\csname @qnnum+3\endcsname
  {{\t@ghead\advance\tagnumber by 3\relax\number\tagnumber}}

\def\equationfile{%
  \@qnfiletrue\immediate\openout\eqnfile=\jobname.eqn%
  \def\write@qn##1{\if@qnfile\immediate\write\eqnfile{##1}\fi}
  \def\writenew@qn##1{\if@qnfile\immediate\write\eqnfile
    {\noexpand\tag{##1} = (\t@ghead\number\tagnumber)}\fi}
}

\def\callall#1{\xdef#1##1{#1{\noexpand\call{##1}}}}
\def\call#1{\each@rg\callr@nge{#1}}

\def\each@rg#1#2{{\let\thecsname=#1\expandafter\first@rg#2,\end,}}
\def\first@rg#1,{\thecsname{#1}\apply@rg}
\def\apply@rg#1,{\ifx\end#1\let\next=\relax%
\else,\thecsname{#1}\let\next=\apply@rg\fi\next}

\def\callr@nge#1{\calldor@nge#1-\end-}
\def\callr@ngeat#1\end-{#1}
\def\calldor@nge#1-#2-{\ifx\end#2\@qneatspace#1 %
  \else\calll@@p{#1}{#2}\callr@ngeat\fi}
\def\calll@@p#1#2{\ifnum#1>#2{\
	@rrwrite{Equation range #1-#2\space is bad.}
\errhelp{If you call a series of equations
 by the notation M-N, then M and N must be integers, and N
 must be greater than or equal to M.}}\else%
 {\count0=#1\count1=#2\advance\count1 by1\relax
  \expandafter\@qncall\the\count0,%
  \loop\advance\count0 by1\relax%
    \ifnum\count0<\count1,\expandafter\@qncall\the\count0,%
  \repeat}\fi}

\def\@qneatspace#1#2 {\@qncall#1#2,}
\def\@qncall#1,{\ifunc@lled{#1}{\def\next{#1}\ifx\next\empty\else
  \w@rnwrite{Equation number \noexpand\(>>#1<<)
  has not been defined yet.}
  >>#1<<\fi}\else\csname @qnnum#1\endcsname\fi}

\let\eqnono=\eqno
\def\eqno(#1){\tag#1}
\def\tag#1$${\eqnono(\displayt@g#1 )$$}

\def\aligntag#1\endaligntag
  $${\gdef\tag##1\\{&(##1 )\cr}\eqalignno{#1\\}$$
  \gdef\tag##1$${\eqnono(\displayt@g##1 )$$}}

\def\eqalignno#1{\displ@y \tabskip\centering
  \halign to\displaywidth{\hfil$\displaystyle{##}$\tabskip\z@skip
    &$\displaystyle{{}##}$\hfil\tabskip\centering
    &\llap{$\displayt@gpar##$}\tabskip\z@skip\crcr
    #1\crcr}}

\def\displayt@gpar(#1){(\displayt@g#1 )}

\def\displayt@g#1 {\rm\ifunc@lled{#1}\global\advance\tagnumber by1
      {\def\next{#1}\ifx\next\empty\else\expandafter
      \xdef\csname @qnnum#1\endcsname{\t@ghead\number\tagnumber}\fi}
  \writenew@qn{#1}\t@ghead\number\tagnumber\else
    {\edef\next{\t@ghead\number\tagnumber}%
    \expandafter\ifx\csname @qnnum#1\endcsname\next\else
    \w@rnwrite{Equation \noexpand\tag{#1} is a duplicate number.}\fi}
  \csname @qnnum#1\endcsname\fi}

\def\ifunc@lled#1{\expandafter\ifx\csname @qnnum#1\endcsname\relax}

\let\@qnend=\end\gdef\end{\if@qnfile
\immediate\write16{Equation numbers written on
  []\jobname.EQN.}\fi\@qnend}

\catcode`@=12


\catcode`@=11
\newcount\r@fcount \r@fcount=0
\newcount\r@fcurr
\immediate\newwrite\reffile
\newif\ifr@ffile\r@ffilefalse
\def\w@rnwrite#1{\ifr@ffile\immediate\write\reffile{#1}
  \fi\message{#1}}

\def\writer@f#1>>{}
\def\referencefile{
  \r@ffiletrue\immediate\openout\reffile=\jobname.ref%
  \def\writer@f##1>>{\ifr@ffile\immediate\write\reffile%
    {\noexpand\refis{##1} = \csname r@fnum##1\endcsname = %
     \expandafter\expandafter\expandafter\strip@t\expandafter%
     \meaning\csname r@ftext\csname r@fnum##1\endcsname
     \endcsname}\fi}%
  \def\strip@t##1>>{}}

\def\citeall#1{\xdef#1##1{#1{\noexpand\cite{##1}}}}
\def\cite#1{\each@rg\citer@nge{#1}}

\def\each@rg#1#2{{\let\thecsname=#1\expandafter\first@rg#2,\end,}}
\def\first@rg#1,{\thecsname{#1}\apply@rg}
\def\apply@rg#1,{\ifx\end#1\let\next=\relax%
\else,\thecsname{#1}\let\next=\apply@rg\fi\next}%

\def\citer@nge#1{\citedor@nge#1-\end-}
\def\citer@ngeat#1\end-{#1}
\def\citedor@nge#1-#2-{\ifx\end#2\r@featspace#1
  \else\citel@@p{#1}{#2}\citer@ngeat\fi}
\def\citel@@p#1#2{\ifnum#1>#2{
	\errmessage{Reference range #1-#2\space is bad.}%
    \errhelp{If you cite a series of references
    by the notation M-N, then M and  N must be integers,
    and N must be greater than or equal to M.}}\else%
 {\count0=#1\count1=#2\advance\count1 by1\relax
  \expandafter\r@fcite\the\count0,%
  \loop\advance\count0 by1\relax
    \ifnum\count0<\count1,\expandafter\r@fcite\the\count0,%
  \repeat}\fi}

\def\r@featspace#1#2 {\r@fcite#1#2,}
\def\r@fcite#1,{\ifuncit@d{#1}%
    \newr@f{#1}%
    \expandafter\gdef\csname r@ftext\number\r@fcount\endcsname%
                     {\message{Reference #1 to be supplied.}%
                      \writer@f#1>>#1 to be supplied.\par}%
 \fi%
 \csname r@fnum#1\endcsname}
\def\ifuncit@d#1{\expandafter\ifx\csname r@fnum#1\endcsname\relax}%
\def\newr@f#1{\global\advance\r@fcount by1%
    \expandafter\xdef\csname r@fnum#1\endcsname{\number\r@fcount}}

\let\r@fis=\refis
\def\refis#1#2#3\par{\ifuncit@d{#1}%
   \newr@f{#1}%
   \w@rnwrite{Reference #1=\number
   \r@fcount\space is not cited up to now.}\fi%
  \expandafter\gdef\csname r@ftext\csname r@fnum#1\endcsname
  \endcsname%
  {\writer@f#1>>#2#3\par}}

\def\ignoreuncited{
   \def\refis##1##2##3\par{\ifuncit@d{##1}%
     \else\expandafter\gdef\csname r@ftext\csname r@fnum##1
   \endcsname\endcsname%
     {\writer@f##1>>##2##3\par}\fi}}

\def\r@ferr{\endreferences\errmessage{I was expecting to see
\noexpand\endreferences before now;  I have inserted it here.}}
\let\r@ferences=\references
\def\references{\r@ferences\def\endmode{\r@ferr\par\endgroup}}

\let\endr@ferences=\endreferences
\def\endreferences{\r@fcurr=0%
  {\loop\ifnum\r@fcurr<\r@fcount%
    \advance\r@fcurr by 1\relax\expandafter\r@fis
    \expandafter{\number\r@fcurr}%
    \csname r@ftext\number\r@fcurr\endcsname%
  \repeat}\gdef\r@ferr{}\endr@ferences}


\let\r@fend=\endpaper\gdef\endpaper{\ifr@ffile
\immediate\write16{Cross References written on
  []\jobname.REF.}\fi\r@fend}

\catcode`@=12

\citeall\refto		
\citeall\ref		%
\citeall\Ref		%



\def\undertext#1{$\underline{\hbox{#1}}$}
\def\frac#1#2{{#1 \over #2}}
\def\12{{1\over2}}
\def\square{\kern1pt\vbox{\hrule height 1.2pt
\hbox{\vrule width 1.2pt\hskip 3pt
   \vbox{\vskip 6pt}\hskip 3pt
   \vrule width 0.6pt}\hrule height 0.6pt}\kern1pt}

\def\ar{ \longrightarrow }
\def\dar{\buildrel \partial_*\over \longrightarrow }
\def\aar{\buildrel \alpha_*\over \longrightarrow }
\def\btar{\buildrel \beta_*\over \longrightarrow }


\baselineskip=\normalbaselineskip \multiply\baselineskip by 2


\null\vskip 3pt plus 0.2fill
   \beginlinemode \baselineskip=\normalbaselineskip
\multiply\baselineskip by 2 \raggedcenter \bf

Generalized Sums over Histories for Quantum Gravity
I. Smooth Conifolds

\vskip 3pt plus 0.2fill \beginlinemode
  \baselineskip=\normalbaselineskip \raggedcenter\sc
        Kristin Schleich

\vskip 3pt plus 0.2fill \beginlinemode
  \baselineskip=\normalbaselineskip \raggedcenter\sc
        Donald M. Witt

\vskip 3pt plus 0.1fill \beginlinemode
   \baselineskip=\normalbaselineskip
  \multiply\baselineskip by 3 \divide\baselineskip by 2
  \raggedcenter \sl

Department of Physics
University of British Columbia
Vancouver, British Columbia V6T 1Z1

\vskip 3pt plus 0.3fill \beginparmode
\baselineskip=\normalbaselineskip
  \multiply\baselineskip by 3 \divide\baselineskip by 2
\noindent ABSTRACT:
Quantum amplitudes for Euclidean gravity constructed by sums over
compact manifold histories are a natural arena for the study of
topological effects.  Such Euclidean functional integrals in four
dimensions include histories for all boundary topologies.  However, a
semiclassical evaluation of the integral will yield a semiclassical
amplitude for only a small set of these boundaries. Moreover, there are
sequences of manifold histories in the space of histories that approach
a stationary point of the Einstein action but do not yield a
semiclassical amplitude; this occurs because the stationary point is
not a compact Einstein manifold. Thus the restriction to manifold
histories in the Euclidean functional integral eliminates semiclassical
amplitudes for certain boundaries even though there is a stationary
point for that boundary.  In order to incorporate the contributions
from such semiclassical histories, this paper proposes to generalize
the histories included in Euclidean functional integrals to a more
general
set of compact topological spaces. This new set of spaces, called
conifolds, includes the nonmanifold stationary points; additionally, it
can be proven that sequences of approximately Einstein manifolds  and
sequences of approximately Einstein conifolds both converge to Einstein
conifolds. Consequently, generalized Euclidean functional integrals
based on these conifold histories yield semiclassical amplitudes for
sequences of both manifold and conifold histories that approach a
stationary point of the Einstein action. Therefore sums over conifold
histories provide a useful and self-consistent starting point for
further study of topological effects in quantum gravity.

\vfill\eject
\beginparmode

\head{1.~Introduction}
\taghead{1.}

An interesting property of our observed universe is, that on scales
ranging from fermis to parsecs, its spatial topology is Euclidean.
This fact is not explained  by the dynamics of classical relativity;
classically all three manifolds admit initial data satisfying the
constraints and this initial data has a finite evolution that produces
a classical spacetime.\refto{dw} Moreover, the spatial topology of the
initial 3-manifold cannot change under evolution.\refto{geroch} Though
classically not allowed, topology change may occur when the quantum
mechanics of gravity is considered.  At distances near the planck
scale, one expects that metric fluctuations will become important and
potentially lead to degeneracies in the metric; such degeneracies can
be argued to signal topology change.  This leads naturally to the
question, what does quantum mechanics of gravity predict for the
topology of our universe?

A formulation of quantum mechanics especially suited to addressing this
question is that given by Feynman's sum over histories.  In particular,
Euclidean sums over histories weighted by the Einstein action provide a
versatile method of constructing amplitudes and states that very
naturally incorporates histories
 corresponding to different topologies.\refto{h1}  A quantum amplitude
 is constructed by summing over all physically distinct histories  that
satisfy the appropriate boundary conditions weighted by the Euclidean
action.  In Euclidean gravity, such a history consists of both a
manifold $M^n$ and a Euclidean metric $g$ on the manifold.  In terms of
such histories, an amplitude for topology change between a  set of
boundary manifolds $\Sigma^{n-1}$  is heuristically $$G[\Sigma^{n-1},
h] = \sum_{(M^n,g)} \  \exp(-I[g])\eqno(topchange)$$ where the sum is
over all physically distinct histories $(M^n,g)$ that satisfy the
appropriate conditions and that have the correct induced metric $h$ on
each boundary. Such a sum over histories will produce an amplitude for
topology change between boundary manifolds in the same cobordism class.
Additionally, it naturally incorporates contributions from histories of
different topologies.  Thus, Euclidean sums over histories provide a
method of incorporating topology into the quantum mechanics of
gravity.
 Indeed, they have frequently been used as a starting point for
 studying the qualitative effects of quantum gravity and topology
change. In fact many interesting investigations of the quantum
mechanics of gravity have been carried out in certain specialized
forms; for example by evaluating expressions such as \(topchange) in
semiclassical approximation in terms of certain known Euclidean
instantons\refto{h1,genlist}
 or in minisuperspace models in which the  histories  summed over are
restricted to a limited set of metrics on a fixed
manifold.\refto{hh,jb} However, in order to address the full
consequences of topology on the quantum mechanics of gravity, the sum
over histories with all topologies, not those in a restricted set
really must be considered. It turns out that two important issues arise
when carefully considering the precise meaning of the heuristic
expression \(topchange); what are the allowed topologies of the
histories that should be included in a more precise sum over histories
and how does one formulate in implementable terms a sum over histories
that have different topologies.

The first issue may seem to be a moot point given that the formal
description of a history to be summed over in \(topchange) is one with
the topology of a smooth manifold. However, this is not the case;
it is well known that formal
descriptions of histories  are based on the classical configurations
of the theory and they do not necessarily correspond to the precise
mathematical definition of the space of histories needed in order to
make Euclidean sums over histories  both well defined and yield the
correct quantum mechanics. For example, the formal description of a
history for a single particle in Euclidean mechanics is a smooth path.
However, it is well known that the paths summed over in a Euclidean
functional integral to produce an amplitude include nondifferentiable
paths.  The contribution of these paths is important; indeed it is well
known that the differentiable ones form a set of measure zero.
Similarly,  analogous nondifferentiable field configurations occur in
functional integrals for field theories.\refto{glimm} Thus one expects
that histories with some sort of suitably nondifferentiable metrics
and matter fields occur in gravity. However, there is an important
difference between histories in field theories and those in Einstein
gravity;  a history in Einstein gravity is specified by
both a  manifold and a  metric and the manifold nature of the
histories is not changed by including nondifferentiable or
distributional metrics in the space of histories.  In light of this, it
is natural to ask whether or not  more general topological spaces
should be included in the sum over histories as the topological analog
of including nondifferentiable paths and if so, what sort of
topological spaces besides manifolds should occur.

The second issue arises even for sets of histories restricted to be
manifolds. In order to proceed from the heuristic expression
\(topchange), it is necessary to provide an implementable description
of how to take the sum over histories. Naively, one imagines
implementing a path integral for gravity by tabulating all distinct
manifolds with the given boundary $\Sigma^{n-1}$, calculating the sum
over some appropriately defined space
of all physically distinct metrics on these manifolds and finally
summing over the contributions for each distinct manifold in the
tabulation weighted perhaps by some phase factor. However, no matter
how reasonable it sounds, such a scheme is not generally possible
because of the first step. There is no way to tabulate all physically
distinct manifolds in dimension $n\ge 4$. Even worse, it is an open
problem to find a method to determine if a space is a manifold in four
dimensions and it is proven that there there is no way to do so for
$n\ge 5$.
Moreover, it is completely independent of such other issues in the
concrete implementation of Euclidean integrals for gravity such as
conformal rotation and the perturbative nonrenormalizability of the
theory.

Finally, the first issue of what topological spaces should be included
as histories is strongly coupled to this problem of finding an
implementable method of summing over distinct spaces; changing the set
of spaces summed over will change the properties of the sum. Thus, as
first suggested by Hartle,\refto{unruly} the inclusion of more general
topological spaces in the sum over histories may provide an avenue
towards finding an expression of the form of \(topchange) that is
better defined.
Thus the twin issues of what set of topological spaces should be
included as histories and of defining a sum over topologies  for
Euclidean
gravitational integrals is as much at hand as the calculation of
possible effects from topology and topology change. These issues will
be examined in a two part paper; Part I will concentrate on the
question of what set of topological spaces should be included as
histories. Part II (Ref. [\cite{II}]) will discuss the issues involved
in  defining a sum over topologies for both manifolds and the set of
more topological spaces, conifolds, proposed in Part I.

The starting point  toward finding a candidate for a more general set
of topological spaces than manifolds is to examine the properties of
the formal expression \(topchange) in terms of semiclassical
approximation. Even though \(topchange) is not fully defined,
semiclassical approximations to it are computable as they are closely
related to the classical solution space of the theory.  The
semiclassical evaluation of a Euclidean sum over histories such as
\(topchange) is constructed from the solution of the Einstein equations
$g$ on a manifold $M^n$ that has boundary $\Sigma^{n-1}$ with the
appropriate induced boundary data $h$. However, although the expression
\(topchange) includes histories for any  boundary $\Sigma^{n-1}$
cobordant to a (n-1)-sphere, there may not be a stationary history for
that boundary.  One can demonstrate that the requirement that the
Euclidean Einstein equations be satisfied on a compact manifold
eliminates a large set of possible boundary manifolds.  Therefore in
the semiclassical limit, transition amplitudes constructed from
Euclidean sums over histories  make strong predictions about the
allowed spatial topology of the universe.

Finding that most boundary manifolds are suppressed in the
semiclassical limit might be considered  a positive result. However, as
discussed in detail in this paper,
further investigation shows that for certain $(\Sigma^{n-1},h)$ that
do not have a semiclassical amplitude, there is a set of smooth
histories consisting of compact manifolds with metrics that approach a
stationary point of the Einstein equations. Nonetheless, there is no
limiting Einstein manifold; the topology of the compact limit space
exhibits nonmanifold points. Therefore a semiclassical evaluation of
\(topchange) does not yield a semiclassical amplitude, precisely
because of the restriction to manifolds. On the other hand,  it is
reasonable to expect that a semiclassical evaluation of a Euclidean
integral that contains a set of histories that approach a stationary
point  yields a semiclassical amplitude.  Thus this feature of the
semiclassical approximation to Euclidean sums over histories for
gravity indicates that their formulation should be generalized: the sum
over histories should include histories corresponding to more general
topological spaces than manifolds.

Given these results, it is natural to investigate the properties and
consequences of extending the sum over  manifolds in Euclidean sums
over histories for gravity to a sum over a more general set of
topological spaces. This paper will concentrate on the properties of a
particular set of such topological spaces, conifolds,
whose definition is motivated by the study of the semiclassical
approximation.
Section 2  will discuss  topological aspects of the manifold histories
used in the formulation of Euclidean sums over histories for Einstein
gravity in terms of the explicit example of the Hartle-Hawking
functional integral for the initial state of the universe.  It will be
manifestly apparent that the general  properties of this particular
Euclidean functional integral are common to all such amplitudes.
Section 3 will discuss in detail why the Euclidean functional
integrals do  not provide semiclassical wavefunctions for all boundary
topologies. As an illustrative example, the case of the Hartle-Hawking
wavefunction for $RP^3$ with round metric will be studied. Section 4
will be devoted to a discussion of the properties desired in a more
general set of topological spaces in order to define and implement a
generalized sum over histories, namely that the sum over histories can
actually be formulated for the set. Section 5 will propose a new set of
topological spaces that satisfy the criteria of section 4; these spaces
will be called conifolds.
Both their topology and geometry will be defined, examples of
conifolds will be presented and various useful results for both
understanding conifolds and using them will be derived.  Section 6
will outline the proof that sequences of approximately stationary
histories converge to a Einstein conifold. It will discuss the
implications this set has in the semiclassical evaluation of Euclidean
functional integrals and propose this set as suitable generalized
histories.

The issue of defining the sum over these topological spaces will be
discussed in Part II, reference [\cite{II}]. Indeed, it will be seen
that conifolds can be described by a simple algorithm in four or fewer
dimensions. Thus certain problems with the sum over histories present
in the case of manifolds will be avoided.
Additionally it will be seen that the simplicial version of conifolds
provide a useful method of formulating discrete forms of \(topchange)
using Regge calculus suitable for the computation of topological
effects.

\head{2.~The Space of Histories}
\taghead{2.}

It is useful to begin by discussing the elements entering into a
heuristic expression such as \(topchange) in terms of an illustrative
example; the Hartle-Hawking wavefunction for the initial state of the
universe.\refto{hh,h2}
The configuration space for the Hartle-Hawking wavefunction is the
space of all closed smooth (n-1)-manifolds $\Sigma^{n-1}$ with three
metrics $h$. These closed (n-1)-manifolds may consist of more than one
disconnected component. Then the Euclidean sum over histories defining
the Hartle-Hawking  state for the case of positive cosmological
constant $\Lambda$ is given by
$$\eqalignno{\Psi[\Sigma^{n-1}, h] &=
\sum_{M^n}\int Dg  \exp\biggl(-I[g] \biggr)\cr I[g] &=- \frac 1{16\pi
G} \int_{M^n}  (R-2\Lambda) d\mu(g)
- \frac 1{8\pi G}\int_{\Sigma^{n-1}}  K d\mu(h) &(hhgstate)\cr}$$
where $d\mu$ is the covariant volume element corresponding to the
indicated metrics.  The boundary conditions that specifically determine
this state are given in terms of the histories included in the sum;
formally, these histories consist of physically distinct metrics $g$ on
compact manifolds $M^n$ that have the correct induced metric $h$ on the
boundary $\Sigma^{n-1}$. Note that, as typically done in the
formulation of Euclidean sums over histories for gravity, the fact that
the set of all compact smooth n-manifolds is countable has been
utilized to write the sum over histories in \(hhgstate) in terms of a
sum over distinct manifolds $M^n$ and a functional integration over
physically distinct metrics on each distinct manifold $M^n$.

This formal description of the histories is the starting point for a
concrete definition of the space of histories and the measure of the
path integral. A complete and detailed specification of this space and
measure is unknown as it is equivalent to demonstrating the existence
of the quantum theory. Indeed a properly defined quantum theory of
Einstein gravity may not exist given its well known problems such as
perturbative nonrenormalizability. However, it is possible to discuss
the topological aspects of the space and measure  as this information
is encoded into the classical  histories. These aspects are important
as they enter into both the semiclassical evaluation of the theory and
discrete approximations of the Hartle-Hawking integral. Thus these
topological aspects are directly relevant to the qualitative study of
Einstein gravity. Moreover, as the topology is
not coupled to the metric, one anticipates that the topological
properties of expressions such as \(hhgstate) are directly relevant to
other theories that include a sum over topologies such as theories that
include gravity or topological field theories.

It is useful to begin by giving the mathematical definitions
corresponding to the formal description of the histories; such
histories will be called Riemannian histories to clearly indicate their
correspondence with classical Riemannian geometry.

\subhead{\undertext{2.1 Riemannian Histories}}

A Riemannian history consists of two quantities;  a metrizable space
corresponding to a smooth manifold and a Riemannian metric on that
manifold. A metrizable topological space is one for which the open sets
of the space can be defined in terms of a distance function.\refto{kn}
A distance function is a real valued symmetric function for which 1)
given any two points $x,y$ in the space, $d(x,y)=d(y,x) \geq 0$ and
$d(x,y)=0$ if and only if $x=y$, 2) the triangle inequality holds,
$d(x,y) + d(y,z) \geq d(x,z)$. Then
\proclaim Definition {(2.1)}.  A
metrizable space $M^n$ is a  smooth manifold if it satisfies the
following conditions:
\item{ 1)} Every point has a neighborhood
$U_\alpha$ which is homeomorphic to an open subset of ${\bf R}^n$ via a
mapping $\phi_\alpha:U_\alpha \to {\bf R}^n$.
\item{ 2)} Given any two
neighborhoods with nonempty intersection, then the mapping $$\phi_\beta
\phi_\alpha^{-1}: \phi_\alpha(U_\alpha \cap U_\beta) \to
\phi_\beta(U_\alpha \cap U_\beta)$$ is a smooth mapping between subsets
of ${\bf R}^n$.\par
\noindent The set $\{(U_\alpha,\phi_\alpha)\}$ is
called an atlas of the manifold; the set $\{U_\alpha\}$ is the cover.
A manifold  for which every cover can be reduced to a finite subcover
is called a {\it compact manifold}.  A manifold that satisfies the
first condition but not the second is called a {\it topological
manifold}. The second condition requires the existence of a smooth
structure on the manifold.  The smooth structure of a manifold is given
in terms of its atlas.  The equivalence of two smooth structures is
thus a question of whether or not two manifolds that are homeomorphic
are actually diffeomorphic; that is given that there is a homeomorphism
$h:M^n\to M'^n$, is the homeomorphism a differentiable, invertible map
whose inverse is also differentiable.  This question is not an issue in
two and three dimensions; it can be proven that all smooth structures
on a given closed connected manifold are equivalent in two and three
dimensions. Indeed, all topological closed manifolds have a unique
smooth structure in two and three dimensions.  In four or more
dimensions, there are topological manifolds that do not admit any
smooth structure;\refto{freedman} the example of $|| E8||$ illustrates
the deduction of this result. Additionally, it can be demonstrated that
there are compact 4-manifolds which have more than one smooth
structure; in fact there exist compact manifolds for which there are a
countably infinite number of smooth structures. For example, the
4-manifold $CP^2\# 9(-CP^2)$, a connected sum of complex projective
space with nine copies of itself with the opposite orientation, has a
countably infinite number.\refto{freedman} Even more interesting is the
result that the number of different smooth structures is uncountable
for certain open manifolds. In particular, ${\bf R}^4$ and ${\bf
R}\times S^3$ both have an uncountable number of distinct smooth
structures.\refto{freedman} In five or more dimensions, the number of
inequivalent smooth structures on compact manifolds is
finite.\refto{toy} Even open manifolds have a finite number of smooth
structures in this case provided that their homology groups are
finitely generated.

However, even though topological manifolds occur in 4 or more
dimensions, it is not necessary to consider them. A smooth structure is
necessary for physical reasons; derivatives of fields can only be
defined on manifolds with smooth structures and such quantities are
fundamental in the discussion of physical theories. Therefore
topological manifolds that are not smooth are not physically
interesting. Finally, one can prove that any $C^1$ atlas is $C^1$
diffeomorphic to a smooth atlas.\refto{diffgeom} Therefore, without
loss of generality, it is sufficient to assume that
the atlas is smooth in the definition of a Riemannian history as done
above.

The definition of a {\it smooth manifold with boundary} differs from
that of a smooth manifold by replacing condition 1) by the requirement
that every point has a neighborhood $U_\alpha$ which is homeomorphic to
an open subset of the upper half space, ${\bf R}^n_+$. The case of
smooth manifolds without boundary is contained in this definition as
open subsets of ${\bf R}^n$ are open subsets of the upper half space.
The boundary of the manifold is given by the set of points that are
mapped to the boundary of the upper half space. From this definition it
follows that the boundary of a smooth n-manifold is a (n-1)-manifold
without boundary. It is important to note that the boundary of a smooth
n-manifold is a topological invariant. Finally,  a compact manifold
without boundary is called a {\it closed manifold}.

The geometry of a Riemannian history is carried by a metric $g$. Smooth
metrics can be found on all smooth manifolds $M^n$ with a smooth
atlas.
 However, it is clear that the set of smooth metrics is too restrictive
as many physically interesting spaces have $C^k$ metrics.  Thus  the
geometry should include $C^k$ metrics with some appropriate choice of
$k$. Given that the action should be defined for a Riemannian history, the
metric $g$  should be at least $C^2$ so that the scalar curvature of
the manifold will be a well defined function. In the case of $M^n$ with
boundary, the metric is restricted by the requirement that it induce
the correct specified metric $h$ on the compact boundary
$\Sigma^{n-1}$. Of course, the degree of differentiability of a given
$h$ will constrain the differentiability of $g$.

Thus the previous paragraphs provide the mathematical foundation for
the formal description of the histories included in the Hartle-Hawking
integral:
A {\it Riemannian history} is a pair  $(M^n,g)$ where $M^n$ is a
smooth compact manifold and $g$ is at least a  $C^2$ metric with the
specified induced metric $h$ on the boundary $\Sigma^{n-1}$.
It is important to stress that the definition of a smooth manifold
does not in any way require the presence of a smooth metric even though
a smooth metric on a manifold can be used to construct the
neighborhoods and maps in Def.(2.1). This point is sometimes overlooked
because of this strong connection between geometry and topology for
Riemannian manifolds. However, it is especially important to keep it in
mind when working with Euclidean functional integrals for gravity as
one should anticipate that all histories will not be classical
Riemannian manifolds. Furthermore, care in separating the issues of
topology and metric can be invaluable in resolving certain confusions
that may arise regarding what constitutes a Riemannian history for the
Hartle-Hawking wavefunction.  For example, an n-ball is an allowed
history with boundary $S^{n-1}$. However, an n-ball minus a small
q-ball around one of its interior points where $q<n-1$ is not; this
manifold is not compact.  Even so, its boundary is still $S^{n-1}$; no
points in the neighborhood of the excised q-ball are mapped to the
boundary of ${\bf R}^n_+$. Note that this example holds even for the
case of a point. This example is obvious when phrased in terms of the
topology; however, the issue can become less clear if one one does not
isolate the topology from the metric.  It is apparent that a careful
recall of these mathematical definitions is useful and necessary for
deciding if a given history is to be included in the Hartle-Hawking
integral.

Although the histories in the Euclidean integrals are formally
Riemannian histories, it is important to remember that the correct
space of histories for Einstein gravity very likely includes not only
Riemannian histories but more general histories.\refto{glimm} For
example,  in Euclidean functional integrals in quantum mechanics,
recall
that a  classical history for a one dimensional free particle is a
differentiable path $x(t)$. However, it is well known that the space of
histories for the path integral for a transition amplitude
$G(a,t;b,t')$ is the space of all continuous paths between the
endpoints $a,b$. This space is much larger than the set of all
differentiable paths between the endpoints.
 One finds that it is necessary to use this larger space space in order
for there to be a well defined measure, the conditional Wiener measure.
This measure replaces both the formal sum over paths and the weighting
by the classical Euclidean action in the path integral; thus it is not
necessary that the classical action by itself be well defined on all
paths in the space. The conditional Wiener measure has  the appropriate
properties such that integration over the space of continuous paths
yields the correct quantum mechanical amplitudes for this system.
Similarly,
the appropriate space of histories for the functional integral for a
free scalar field includes distributional  fields $\phi(x)$ such that
$\int_{{\bf R}^n} f(x)\phi(x)dx$ is finite where $f$ is a smooth test
function with compact support. Of course, the  correct space of
histories and measure for a general interacting field theories are
unknown; such a formulation is equivalent to a solution of the full
quantum field theory and is therefore a highly nontrivial matter.
However it is expected that the space of histories for any general
field theory is larger than the set of classical histories of that
theory.  Therefore,  one anticipates that the space of histories in the
Hartle-Hawking wavefunction will include not only all $C^2$ metrics $g$
but also less regular, distributional metrics defined in a manner
similar to that for distributional fields.  Again, having a well
defined action $I(g)$ for a distributional history  is not directly
relevant for the issue of defining the space of histories.  As for
quantum mechanics, one anticipates that the measure on this space
replaces both the integration over physically distinct metrics
and the weighting factor of the action in \(hhgstate).

Given that such distributional metrics are likely to be included in the
space of histories, it is important to stress that a distributional
metric does not imply that the underlying manifold is somehow singular.
The smooth structure on the manifold is used in the definition of the
space of distributional fields or metrics through the integration
against a smooth test field. What is true is that the classical
correspondence of a smooth metric with the topology of the manifold no
longer holds for distributional fields.
This is illustrated by the familiar example of the one parameter
minisuperspace model in which the radius of the (n-1)-sphere is the
only degree of freedom.\refto{hh} The histories for the Hartle-Hawking
integral for this model consist of the set of spherically symmetric
metrics on the n-ball. As there is one degree of freedom, the integral
is of the same form as one for a quantum system with one degree of
freedom.  Thus, one anticipates that the space of histories will
include continuous paths. Continuous spherically symmetric histories
can written in gauge fixed form as $ds^2 = d\tau^2 + a^2(\tau) d
\Omega^2$ where $a$ is a continuous function on the interval $[0,1]$
that vanishes at $0$ and $d\Omega^2$ is the round (n-1)-sphere metric.
Note that even though the metrics are  continuous not differentiable,
the topology of the underlying manifold is still that of  a smooth
n-ball. Also, even though the metric may not be differentiable at the
point $\tau=0$, this point is not a boundary of the n-ball by
definition of boundary. Therefore a minisuperspace path integral based
on this set of metrics does not change the topology of the n-ball or
have any boundaries other than those supplied as boundary conditions,
even though the connection between the distance function, topology and
smooth structure does not hold for some of the metrics in the set.
Similarly, the inclusion of distributional metrics in the space of
histories does not change the topological aspects of this space. It
again emphasizes the point that
the topology of a manifold is specified independently of its metric.
Additionally, it implies that  the topological aspects of defining the
space of physically distinct histories can be discussed in terms of the
Riemannian histories alone.

Given that the topology of a Riemannian history is any smooth compact
manifold $M^n$ which has as its boundary the closed manifold
$\Sigma^{n-1}$, an immediate question is whether or not the set of
allowed $M^n$ is empty for a  given closed  $\Sigma^{n-1}$. The answer
to this question depends on the dimension and topology of the boundary
manifold. In two dimensions, the set of $M^2$ is not empty for any
closed boundary manifold.  The closed boundary $\Sigma^1$ simply
consists of the disjoint union of circles. It is easy to see that the
desired  $M^2$ can be constructed from any closed 2-manifold; excise
the interiors of the required number of disjoint discs from the
manifold. In higher dimensions, the answer to this question can be
determined from the cobordism class of the boundary manifold; two
closed  (n-1)-manifolds in the same cobordism class are the boundary of
some compact n-manifold. Consequently, any closed  (n-1)-manifold
cobordant to an (n-1)-sphere is itself the boundary of a compact
n-manifold: A (n-1)-sphere is the boundary of an n-ball and thus the
cobordism between these two boundary manifolds can be capped off at the
(n-1)-sphere to form a compact manifold with the desired boundary. The
set of closed 2-manifolds has two cobordism classes; closed 2-manifolds
with even Euler characteristic and closed 2-manifolds with odd Euler
characteristic. As the Euler characteristic of a 2-sphere is even, all
closed 2-manifolds with even Euler characteristic are the boundaries of
some compact 3-manifold. Therefore the set of allowed histories $M^3$
is not empty for boundary $\Sigma^2$ with even Euler characteristic.
However, closed 2-manifolds with odd Euler characteristic are not
cobordant to the 2-sphere; therefore, there are no allowed histories in
the Hartle-Hawking wavefunction for these boundaries.  All closed
3-manifolds are in the same cobordism class; therefore, the set of
allowed histories $M^4$ for any given $\Sigma^3$ is not empty. A
similar analysis can be done in 5 or more dimensions to determine
whether or not the set of n-manifolds with a specified boundary
(n-1)-manifold is empty.

Indeed, the cobordism properties of manifolds imply that it is
inconsistent to restrict the topology of the n-manifolds included in
the set of allowed histories without strong restrictions on the allowed
boundary (n-1)-manifolds and conversely. For example, consider
restricting the allowed boundary manifolds to be  3-spheres. As any
arbitrary  3-manifold is cobordant to $S^3$, an arbitrary history in
Hartle-Hawking wavefunction for two  $S^3$ boundaries will include an
intermediate hypersurface with arbitrary topology $\Sigma^3$.
Therefore, the set of histories with $S^3$ boundary can be used to
generate a set of histories with arbitrary boundary $\Sigma^3$. Thus if
the set of all $M^4$ cobordant to $S^3$ are allowed histories, then the
configuration space of the wavefunction must include all $\Sigma^{3}$;
conversely if arbitrary boundary 3-manifolds are allowed, the allowed
histories must include all cobordisms. This argument can easily be
extended to any dimension by discussing everything in terms of
cobordism classes.

If the set of manifolds $M^n$ cobordant to a boundary manifold
$\Sigma^{n-1}$ is not empty, then the next step is to discuss how to
select physically distinct histories for inclusion in the sum in
\(hhgstate).

\subhead{\undertext{2.2 Physically Distinct Histories and the
Problem of Decidability}}

There are two parts to finding the space of physically distinct
histories for integrals such as \(hhgstate); finding the space of
physically distinct metrics on each manifold $M^n$ and finding the set
of physically distinct manifolds $M^n$. The issues involved in finding
the space of physically distinct metrics are familiar from the studies
of gauge theories;
it is well known that the metric $g$ does not uniquely determine the
geometry of the Riemannian history. As Einstein gravity is a
diffeomorphism invariant theory, metrics that are equivalent under a
diffeomorphism of the manifold correspond to the same physical space:
given two histories  $(M^n,g)$ and  $(M^n,g')$  if $g= f^*g'$ under a
diffeomorphism $f:M^n \to M^n$ of the manifold, then the two histories
are equivalent. Note that the diffeomorphism $f$ is not required to be
smooth; thus a smooth metric $g$ may be equivalent to a $C^k$ one. In
order to find a set of physically distinct histories for a given
manifold $M^n$,  it is necessary to restrict to physically distinct
metrics. This can be done formally on each manifold by taking the space
of all suitably defined metrics that are inequivalent under
diffeomorphisms. Of course an implementation of this space will
encounter many of the same difficulties as found in Yang Mills
theories.  For example, additional complications are introduced by the
appearance of $\theta$ sectors for the space of metrics that must be
properly handled.\refto{donandjim} Nonetheless, the issues encountered
are isolated from those relating to the sum over topologies in
\(hhgstate) and thus can be put aside
when addressing topological issues.

The problem of finding the set of physically distinct manifolds $M^n$
is not one encountered in the study of gauge theories as the background
manifold that the gauge fields are defined on is typically fixed.
Thus the question arises as to whether or not  two smooth manifolds
are physically equivalent, that is whether or not they have the same
topology and smooth structure. This question can be separated into
three distinct issues.  The first is whether or not  it is possible to
show that there is a finite method of determining that a given
topological space satisfies the definition of a smooth manifold. The
second is whether or not it is possible to show that the two manifolds
are homeomorphic to each other. The third is whether or not two
manifolds that are homeomorphic  have equivalent smooth structures. The
second issue turns out to be related to the first.

A full and detailed discussion of these three issues will be given in
Part II of this paper as it is necessary to introduce additional
mathematical tools in order to adequately address them.  Essentially, a
finite representation of a smooth manifold is needed for a discussion
of finite methods for determining properties of topological spaces.
However, it is useful to have a brief introduction to
the issues and the results for the purposes of this paper.\refto{IIa}

The first issue is called the algorithmic decidability of n-manifolds:
whether or not there is an algorithmic description of the set of all
n-manifolds. The
algorithmic description of n-manifolds follows the same logic as the
algorithmic description of other things such as flora and fauna. For
example, given the set of all fauna on earth, one can look for an
algorithmic description of a bird that will select out the subset of
all birds on earth. This algorithmic description is a  set of rules
that can be implemented to determine whether or not a given animal in
the set of all fauna is in fact a bird. Such a set of rules can be a
series of questions such as does the animal have feathers, does it have
a beak, does it have wings, and so on. By applying these rules, the
subset of all birds can be selected out from the set of all fauna. Such
an algorithm  must be known to take a finite number of steps in order
to be useful. In the case of birds, it is clear that there exists some
finite set of rules by which this can be done simply because there are
a finite number of fauna on Earth and ipso facto, a set of rules can be
divised to divide this finite set into two, birds and not birds.
Therefore the set of all birds is algorithmically decidable.

Clearly, any finite set is algorithmically decidable. Equally clearly,
any uncountable set is not  algorithmically decidable. Therefore the
question of algorithmic decidability is nontrivial only in the case of
countably infinite sets and there are examples of sets with and of sets
without  algorithmic descriptions. As the set of all compact
n-manifolds is countably infinite, their algorithmic decidability is
nontrivial and it turns out to depend on the dimension n.  As discussed
in Part II, in one, two and three dimensions, the set of all compact
manifolds is algorithmically decidable. In four dimensions, whether or
not there is an algorithmic description of all compact 4-manifolds is
an open problem; the existence of such a description relies on whether
or not there is an algorithm for recognizing a 4-ball. The existence of
such an algorithm relies on solutions to the Poincare conjecture and to
the word problem for the fundamental groups of 3-manifolds, both open
problems in topology. In five or more dimensions, n-manifolds are not
algorithmically decidable; it has been proven that there is no
algorithm for recognizing a n-ball for $n\ge 5$. Thus in five or more
dimensions, there is no algorithmic method of constructing the space of
histories formally summed over in the Hartle-Hawking integral.

The second  issue  is called  the classifiability of n-manifolds; does
there exist a method of determining whether or not a given n-manifold
is homeomorphic to another n-manifold. Again, the classification of
n-manifolds follows the same logic used in classification of other
things. In order to classify birds, for example, one needs a set of
rules  by which it can be determined whether or not a given bird is in
fact a member of a previously identified type or whether it is a
distinct, new bird to be added to the list. This set of rules is based
on a finite algorithm for determining whether or not the given bird is
the same as another specific bird, say a robin. Such an algorithm can
be a series of questions such as is the bird the same length as a
robin, is its breast the same color as a robin's, is its beak the same
shape as a robin's, and so on. If it can be established in a finite
number of steps that a bird is not a robin, then one can go on to the
next bird, say a crow, and use a similar procedure and continue until
the bird is classified as either being a previously identified type or
a distinct new bird.

Again, it is clear that the issue of classifiability is only nontrivial
for countably infinite sets such as n-manifolds. It turns out that
closed connected 1-manifolds are classifiable; the only closed
connected 1-manifold is a circle. Whether or not two closed connected
2-manifolds are in fact the same can be determined by comparing their
orientations and Euler characteristics. Thus closed connected
2-manifolds are classifiable. The classifiability of closed connected
3-manifolds is an open problem; the existence of an algorithm again
depends on several open conjectures in the topology of 3-manifolds.
Finally, it can be proven that closed connected n-manifolds are not
classifiable in four or more dimensions using the unsolvability of the
word problem for finitely presented groups.  It is important to note,
as discussed in more detail in Part II, that this is a problem in
practice, not just in principle; one can find explicit finite
representations of manifolds with explicit undecidable groups as their
fundamental groups.  Thus in four dimensions, there is no algorithm for
providing a list  of physically distinct 4-manifolds necessary for
constructing the space of physically distinct histories.

The third issue is the equivalence of smooth structures on a manifold
$M^n$.  In four or more dimensions, a given closed connected manifold
can have a countable number of smooth structures. These structures are
physically distinct and thus also must be included separately in the
sum over histories.  Again, in order to concretely implement such a sum
one needs a method of deciding whether or not two manifolds $M^n$ and
$M'^n$ that are homeomorphic are also diffeomorphic. However, by the
results of the second issue, one immediately runs into trouble; as
there is no method of classifying manifolds in four or more dimensions
up to homeomorphism, it is clear that there is no finite algorithm for
finding representatives of distinct smooth structures.

These three  issues have several important implications for the
Hartle-Hawking functional integral.  First, they imply that it is not
possible to define the space of physically distinct histories in four
or more dimensions;
there is no way to enumerate all physically distinct compact $M^n$
with boundary $\Sigma^{n-1}$. Second, there is no way to
algorithmically construct a Riemannian history  in five or more
dimensions and a problem with doing so in four dimensions. Therefore,
there is no way to construct the space of histories in these dimensions
let alone the space of physically distinguishable ones. Third,
the formal split between manifolds and metrics in \(hhgstate) ignores
the point that different smooth structures on a given manifold $M^n$ of
four or more dimensions correspond to physically distinct histories and
should be included independently in the sum. As the number is
countable, it can be formally included by replacing the sum over $M^n$
\(hhgstate) with a sum over $(M^n,s)$ where $s$ is an integer labeling
the different smooth structures on $M^n$.  However, though this formal
addition to the measure for the Hartle-Hawking wavefunction is
important, it is also undecidable. Moreover, it in no way affects the
problems  of classifiability and algorithmic decidability.

Thus, the sum over physically distinct manifolds in \(hhgstate) is
formally well defined only in two dimensions; it is clearly not well
defined in  four or more dimensions and may or may not be so in three.
These problems for a concrete formulation of the space of physically
distinct histories and the measure for Hartle-Hawking integral persist
even in terms of a finite approximation as discussed in Part II.
Thus the problems with the topological aspects of the Hartle-Hawking
wavefunction cannot be avoided by simply going to discrete models.
Therefore, if an expression of the form \(hhgstate) is to be replaced
by a well defined meaningful sum over topologies, the fundamental
question of what a physically distinct history is must be addressed.

As first observed by Hartle,\refto{unruly} one way to make such a sum
over topologies well defined is by finding an appropriate set of more
general topological spaces which are algorithmically decidable. Such
generalized spaces can be thought of as the topological analog of
nondifferentiable paths.
Thus it is reasonable to explore the option of summing over more
general topological spaces in order to make expressions such as
\(hhgstate) well defined.  However, this abstract criterion is not
enough; there are many algorithmically decidable spaces. Therefore it
is useful to have a more physical motivation for what more general
topological spaces should be used.
Such motivation can be provided by a study of the semiclassical
approximation.  In semiclassical approximation, only histories which
correspond to the stationary points of the action  contribute to the
evaluation of a functional integral; thus semiclassical approximations
do not require a precise definition of the space of physically distinct
Riemannian histories in order to be carried out. However, as seen in
the next section, such stationary points need not be manifolds. Thus
they provide models for more general topological spaces for inclusion
in the sum over histories.

\head{3.~Euclidean functional integrals in Semiclassical Approximation}
\taghead{3.}

A semiclassical evaluation of the Euclidean functional integral for a
wavefunction of the form \(hhgstate)
involves finding  an appropriately differentiable metric
corresponding to an extremum of the action on some compact manifold
$M^n$. Such a semiclassical history typically consists of a metric that
is Euclidean at small geometries and is Lorentzian at large geometries.
If there is more than one extremum of the action, the semiclassical
approximation will consist of a superposition of the extrema although
in practice often one keeps only the dominant contribution. As a
consequence of the continuity of the extremizing metric, the resulting
transition amplitude will be continuous  with one continuous functional
derivative on the space of three geometries.\refto{continuity}

A now familiar illustration of the semiclassical approximation is
provided by    the case of the Hartle-Hawking wavefunction for  a
boundary  3-sphere with round metric of radius $a_0$.\refto{hh,h2} The
unit 3-sphere metric can be written
$$\eqalignno{d\Omega^2 &= \left( d\chi^2 +
\sin^2\chi   d \theta^2 +\sin^2\chi\sin^2\theta d\phi^2 \right)
&(S3metric) \cr &0\le\chi\le \pi\ \ \ \ \ \ 0\le\theta  \le \pi
\ \ \ \ \ \  0\le  \phi\le 2\pi.&(ranges)\cr}$$
For $Ha_0<1$, a particularly simple extremum of the gravitational
action \(hhgstate) is the  Euclidean de Sitter metric
$$\eqalignno{ ds^2 &= d\tau^2 + a^2(\tau) d\Omega^2\cr a(\tau) &=
\frac 1H\sin(H\tau)&(acl)\cr}$$
where $3H^2=\Lambda$. The topology of this solution is $S^4$; indeed it
is metrically a round 4-sphere. The scale factor $a(\tau)$ explicitly
satisfies the Einstein equations, which reduce to
$$ \frac {\partial_\tau^2 a}{a}
-\left(\frac{\partial_\tau a}{a}\right)^2 + \frac {H^2
}{a^2} =0\eqno(eom)$$
for the metric \(acl). There are two possible positions for the
3-sphere boundary in the Euclidean solution \(acl)  that yield the
correct induced metric on the 3-sphere; they correspond to filling
either less than or more than half the 4-sphere.  According to Hartle
and Hawking, the Euclidean extremum that dominates in the
steepest descents evaluation is that with least action corresponding to
filling less than half the 4-sphere. The wavefunction is thus
$$\eqalignno{\Psi_E({S^3},a_0) &\sim \exp-I^{_-}(a_0)\cr I^{_-}(a_0)
&=  -\frac {1}{3H^2\ell^2} [(1-H^2 a_0^2)^\frac 32 -1].&(eucex)\cr}$$
For $Ha_0>1$, there are no real Euclidean extrema.  Instead, there are
two complex extrema corresponding to the Lorentzian de Sitter solution
$$\eqalignno{ ds^2 &= -d t^2 + a^2(t) d\Omega^2\cr a(t) &=  \frac
1H\cosh(H t).&(lords)\cr}$$
The topology of the Lorentzian de Sitter solution is ${\bf R}\times
S^3$.  The extrema contribute equally to the stationary phase
approximation.  An analysis of the contour of steepest descents leads
to the phase for the semiclassical wavefunction:\refto{hh}
$$\eqalignno{\Psi_L(S^3,a_0) &
\sim \cos(S(a_0)-\frac \pi{ 4})\cr S(a_0)&= \frac 1{3H^2\ell^2}(H^2
a_0^2-1)^{\frac 32}.  &(lorex)\cr}$$
\noindent This simple example concretely illustrates three generic
features about the semiclassical approximation to the Hartle-Hawking
wavefunction and more generally, semiclassical approximations of
gravitational Euclidean functional integrals:

\noindent 1) There may exist regions of the configuration space of all
(n-1)-metrics on (n-1)-manifolds $(\Sigma^{n-1},h)$
for which the extremum of the action is a  Einstein manifold. These
regions correspond to boundaries with diameter smaller than that
given by Myers' Theorem.\refto{kn} The diameter of a closed n-manifold
is defined as the least upper bound of the distance between any two
points; equivalently it is the length of the longest globally
minimizing geodesic between any two points. Then
\proclaim Theorem (Myers) {(3.1)}.  Let $M^n$ be a complete Riemannian
manifold with metric $g$ which satisfies $R_{ab}\ge \kappa^2g_{ab}$ and
$\kappa$ is a nonzero constant. Then $M^n$ is compact with diameter
$d\le \frac \pi \kappa$.\par
This is a  general theorem, but  can of course be applied to Riemannian
manifolds that are solutions of the Euclidean Einstein equations
provided that their Riemannian curvature is manifestly bounded in the
required fashion.  For positive cosmological constant, the Euclidean
Einstein equations reduce to precisely the form required in Thm.(3.1)
with $\kappa^2= \frac 2{n-2}\Lambda$ for $n>2$.  Thus if the diameter
of  $(\Sigma^{n-1},h)$ is less than $\pi/\kappa$, there may be a
Einstein manifold with that boundary.  However, there is no existence
theorem guaranteeing that such an extremum exists for all boundary data
and boundary manifolds.

\noindent 2) There is a region of configuration space $(\Sigma^{n-1},
h)$ for which there is no extremum of \(hhgstate) corresponding to a
real Euclidean solution.  Given  any Einstein manifold with positive
curvature,
Myers' theorem determines its maximum diameter in terms of the
curvature.   Consequently, if the diameter of $\Sigma^{n-1}$ is larger
than this maximum, there is no classical Euclidean solution for this
boundary data.  However, although there are no stationary Euclidean
paths for such  large 3-geometries, there are complex
stationary paths. Examples of such paths are Lorentzian solutions of
the Lorentzian Einstein equations. Initial data which can be evolved to
a  Lorentzian solution to the Einstein equations exist on all
3-manifolds.\refto{dw} For example, suppose $\Sigma^3$ is a hyperbolic
3-manifold, that is one admitting metrics with negative curvature. Pick
an  initial $h$ such that $^3R = - 6H^2$ and let $K^L_{ij}=K_{ij} =
\sqrt{2} H h_{ij}$. This initial data set satisfies the Hamiltonian
constraints for the Lorentzian Einstein equations with positive
cosmological constant, $$\eqalignno{K^{ij}K_{ij} - K^2 -^3R +2\Lambda
&= 0\cr \nabla_i(K^{ij} - Kh^{ij}) &= 0.&(lconstraints)\cr}$$
Therefore, by  standard existence theorems for solutions to the
Einstein equations, it can be evolved for a finite distance. The same
argument can be used show the existence of initial data sets for
3-manifolds admitting positive or zero curvature by appropriate changes
in sign and constants. The action in \(hhgstate) when evaluated for
Lorentzian solutions of the Einstein equations is purely imaginary;
therefore the wavefunctional is oscillatory
in the corresponding regions of $(\Sigma^{n-1},h)$.

\noindent 3) The wavefunction and its functional derivative are
continuous everywhere in the configuration space.  In the $S^3$
example, it can be easily checked that this property holds even at the
point $Ha_0=1$ between the Lorentzian and Euclidean regions.  This
continuity was enforced by the contour of steepest descents but it also
follows directly from basic properties of the Wheeler de Witt
equation.\refto{continuity} In lowest order semiclassical
approximation, it is equivalent to
 matching conditions for the wavefunctionals at certain points in the
 space of all three geometries, i.e. at  boundary three geometries
between Euclidean and Lorentzian wavefunctionals. Thus on each
component of the boundary $\Sigma^{n-1}_b$ with metric $h^b$ between
the Euclidean and Lorentzian wavefunctionals
$$ \frac{\delta
\ }{\delta h_{ij}}\log \Psi_E[\Sigma^{n-1}_b,h^b]  = \frac {\delta
\ }{\delta h_{ij}}\log \Psi_L[\Sigma^{n-1}_b,h^b].\eqno(match)$$
For generic semiclassical wavefunctionals, these boundary conditions
are satisfied for $\pi_L^{ij} =\pi^{ij}_E= 0$ where  $ \pi_L^{ij}
=\frac{\delta S}{\delta h_{ij}}$ and $ \pi^{ij}_E=\frac {\delta
I}{\delta h_{ij}}$ are the classical Lorentzian and Euclidean momenta
respectively. By the standard relation between the extrinsic curvature
and momenta in gravity, this boundary condition implies that the
extrinsic curvature of the boundary vanishes,
$$K^L_{ij} = 0 =K^E_{ij}. \eqno(kmatch)$$
For special cases of the
steepest descents evaluation, the Lorentzian wavefunction or the
Euclidean wavefunction may have values allowing \(match) to be
satisfied for boundaries with nonvanishing extrinsic curvature.
However, one is lead to believe that these special cases are the
exception, and that the generic situation requires \(kmatch) at the
boundary hypersurface.

The semiclassical approximation does not explicitly restrict the
allowed boundary topologies in the Hartle-Hawking wavefunction.
However, its implementation does indeed lead to such restrictions. This
result follows from the three general properties of the semiclassical
approximation to the functional integral discussed above. Most
obviously, the vanishing of the extrinsic curvature requires, from
\(lconstraints), that the scalar curvature of the boundary manifold is
positive definite, $^3R = 2\Lambda$. Thus the boundary manifold must be
one that admits positive curvature.  Such manifolds are $S^3$,
$S^2\times S^1$, $S^3/\Gamma$ where $\Gamma$ is a finite group and
connected sums of these manifolds. This is a small subset of the
countably infinite number of three manifolds as can be demonstrated by
a simple argument using results due to Thurston.\refto{thurston}
Similar results hold for any matter source with positive stress energy.
Thus the boundary conditions \(kmatch) imply that the kinds of topology
change allowed in semiclassical approximation are limited.  However,
semiclassical topology changing amplitudes may not be allowed even for
boundary manifolds in this limited set due to the first and second
properties discussed above.  For example, the Hartle-Hawking proposal
does not yield a semiclassical amplitude for $RP^3$ with round 3-sphere
metric.

The manifold $RP^3$ with  its round metric can be constructed from
$S^3$ with \(S3metric) by identifying antipodal points;
$$\chi =\pi -
\chi;\ \ \theta = \pi - \theta;\  \ \ \ \phi = \phi + \pi
\eqno(idcoord)$$
Thus the unit metric and range of
coordinates on $RP^3$ is
$$\eqalignno{d{\bar \Omega}^2 & =
\left(d\chi^2 + \sin^2\chi d \theta^2 +\sin^2 \chi \sin^2 \theta
d\phi^2\right)&(RP3metric) \cr & 0\le\chi\le \pi \ \ \ \ \ \ 0\le
\theta\le \pi \ \ \ \ \ \ 0\le\phi\le\pi &(idsrange)\cr}$$
differing from the ranges \(ranges) for the $S^3$ round metric by a
factor of $1/2$ for the $\phi$ component. Locally, the manifolds $S^3$
and $RP^3$ have the same metrics; they are only distinguished by their
global properties. Thus, it follows that the equations of motion are
the same for the $RP^3$ case as for the $S^3$ case because they are
local.  Consequently a Lorentzian solution of the Einstein equations
corresponding to $RP^3$ with round metric is identified de Sitter,
$$\eqalignno{ ds^2 &= -dt^2 + a^2(t) d{\bar \Omega}^2\cr a(t) &=  \frac
1H\cosh(Ht)&(lrp3)\cr}$$
with topology ${\bf R}\times RP^3$.  It is a nonsingular Lorentzian
spacetime; it has the same local metric as the $S^3$ de Sitter
spacetime \(lords), although the global properties of the spatial
hypersurfaces of the two spacetimes are   different.
 The Lorentzian extrema of \(hhgstate) for $RP^3$ de Sitter correspond
 to complex stationary points of the Euclidean action.  Thus they would
be expected to  contribute to the stationary phase approximation of the
Hawking Hartle wavefunction for $Ha_0>1$ with action
$$\eqalignno{
\Psi_L({RP^3},a_0) & \sim cos({\bar S}(a_0) +\alpha) \cr {\bar S}(a_0) &=
\frac 1{6H^2\ell^2}(H^2 a_0^2-1)^{\frac 32} &(idlorex)\cr}$$
with the phase $\alpha$ to be determined either by steepest descents
contour or by matching conditions \(match) at $Ha_0=1$. The action
\(idlorex) differs by a factor of $1/2$ from that of the $S^3$ case
\(lorex) due to the difference in volume of the boundary metrics.

The difficulty with constructing a semiclassical wavefunction occurs
when looking for compact manifolds with real Euclidean metrics that
satisfy the Einstein equations and match the boundary data on $RP^3$.
Locally, the equations of motion for explicitly spherically symmetric
solutions  are the same as those for the $S^4$ Euclidean de Sitter
solution \(eom); thus one locally unique solution in a neighborhood of
the initial $RP^3$ with round metric  is
$$\eqalignno{d s^2 &= d\tau^2
+a^2(\tau)d{\bar \Omega}^2\cr a(\tau) &=  \frac
1H\sin(H\tau)&(idacl)\cr}$$
This metric is well defined between two
$RP^3$ hypersurfaces with $a(\tau_0) = a_0$ and $a(\tau_1) = a_1$; the
corresponding range of $\tau$ is $0<  \tau < \frac {2\pi}{H}$. The
manifold of the solution has product topology
${\bf R}\times RP^3$.  The Euclidean action  between the two $RP^3$
hypersurfaces  is well defined
$${\bar I}(a_0,a_1)= - \frac
{1}{6H^2 \ell^2} [(1-H^2 a_0^2)^\frac 32 -(1-H^2 a_1^2)^\frac
32].\eqno(ideucex)$$
However, there is no {\it compact Riemannian manifold} with this
solution as its metric. If  the range of $\tau$ is extended to $\tau=0$
in \(idacl) then indeed $a = 0$ at this point.
The corresponding topological space is compact.
However, the global structure of this compact topological space is not
that of a manifold by Def.(2.1); rather it is a cone over $RP^3$. In
order to discuss this issue, it is necessary to introduce the
definitions of a join, a cone and a suspension.\refto{spanier}
\proclaim Definition {(3.2)}.  Let $U$ and $V$ be topological spaces.
Their join, $U*V$ is the space formed by the cartesian product of $U$,
$V$ and the unit interval $I$ modulo an equivalence relation,
$U*V=(U\times V\times I)/\sim$ where $$(u,v,t) \sim (  u^\prime,
v^\prime, t^\prime) \ \ \cases{ t= t^\prime=0\ \hbox{and}\ u=
u^\prime\cr \hbox{or}\cr   t= t^\prime=1\ \hbox{and} \ v=
v^\prime\cr}.$$\par
\proclaim Definition {(3.3)}.  The cone $C(V)$ is the join of the
topological space $V$ with a point $\{p\}$, $C(V) = V*p$.\par
\proclaim Definition {(3.4)}. The suspension $S(V)$ is the join of the
topological space $V$ with the zero dimensional circle,
$S(V)=V*S^0$.\par
Figure 1 provides an example of a join and Figure 2 that of a
suspension.  Note that the suspension of a topological space is
equivalent to gluing two cones of the space together at their boundary.
For example, the cone of $S^n$ is the (n+1)-ball $B^{n+1}$, the
suspension of $S^n$ is $S^{n+1}$ and it is equivalent to gluing two
$B^{n+1}$ together along their $S^n$ boundary. In general, cones and
suspensions of arbitrary topological spaces $V$ will not produce
manifolds.

The compact topological space formed by the Euclidean solution \(idacl)
for $0\le\tau<\frac {2\pi} H$ is  $C(RP^3)$. This is obvious from
Defs.(3.2-4); the metric \(idacl) is homeomorphic to the cartesian
space $(RP^3, \{p\}, \tau)$ with all points at $\tau = 0$  identified
to a single point. The compact space formed by the Euclidean solution
\(idacl) for $0\le\tau\le \frac {2\pi} H$ is $S(RP^3)$; a
representation of this suspension is given in Figure 3.  The space
$C(RP^3)$ is compact by definition. However,
 it is not a compact manifold.  In order to prove this, note that given
any n-manifold $M^n$ with $n\ge3$, one can readily demonstrate that
$\pi_1(M^n-\{p\}) = \pi_1(M^n)$ for any point $p\in M^n$. This is due
to the fact that in three
 or more dimensions, curves can always be moved around an isolated
 point without ever crossing the point itself. Now assume that
$C(RP^3)$ is a manifold and take the point $p$ to be the apex of the
cone, $\tau = 0$. Note that by construction $C(RP^3) - \{p\} = I\times
RP^3$. Hence $\pi_1(C(RP^3)-\{p\}) = \pi_1(I\times RP^3)) = Z_2$.
However, note that $C(RP^3)$ is contractible  which implies that
$\pi_1(C(RP^3)) = 1$. Therefore, by contradiction, it follows that
$C(RP^3)$ is not a manifold. Consequently, there is no classical
Euclidean solution for $RP^3$ corresponding to the Euclidean de Sitter
solution for $S^3$.  By construction, it follows that there is no
semiclassical Euclidean wavefunction for $RP^3$ with round metric
corresponding  to that given by the contribution of the Euclidean de
Sitter for $S^3$.  Therefore there is no semiclassical approximation to
the Hartle-Hawking wavefunction for a boundary  $RP^3$ with round
metric even though there is a semiclassical Lorentzian wavefunction for
this boundary geometry.

One objection to this conclusion is that this model is too restrictive;
it implicitly assumes that the Euclidean solution is of the form
\(idacl).  However, it can be argued that there is no Einstein manifold
with $RP^3$ boundary with round metric that allows  the continuity
conditions \(match) to be satisfied at $Ha_0=1$. Note that $Ha_0=1$ is
a maximal slice in the Lorentzian identified de Sitter metric \(lrp3).
Consequently, $K^L_{ij}=0$. Therefore, given that the Lorentzian
wavefunction is of the form \(idlorex), the matching conditions
\(match) imply that $K^E_{ij}$ must also vanish identically.  Thus
 the compact Euclidean Einstein manifold with $RP^3$ boundary sought
 has vanishing extrinsic curvature and intrinsic metric \(RP3metric).
 This manifold with boundary can be doubled over at the $K^E_{ij}=0$,
 that is at the maximal
 slice, to form a closed Einstein manifold.  Next, the locally unique
evolution of this maximal slice is the Euclidean solution \(idacl).
Therefore the Weyl tensor vanishes, $C_{abcd}=0$ by explicit
calculation in a neighborhood of the maximal slice. Therefore, the
scalar function given by the square of the Weyl tensor vanishes as
well, $C^{abcd}C_{abcd}=0$, in this neighborhood.
 Now, Einstein manifolds are analytic.\refto{besse} Therefore if a
 function vanishes on an open set of the manifold, it must vanish
everywhere. As $C^{abcd}C_{abcd}$ is positive definite, it follows that
it must vanish everywhere. This implies that $C_{abcd}=0$ everywhere.
Einstein manifolds with vanishing Weyl tensor are space-forms and
 there are two space-forms with constant positive curvature; $S^4$ and
$RP^4$. But $RP^3$ does not divide either of these manifolds into two
manifolds with boundary $RP^3$. A proof of this assertion can be
derived using the Mayer-Vietoris homology sequence for the
decomposition of a topological space; the full argument is given in
appendix A.  Therefore, there are no Einstein manifolds with $RP^3$
boundary with round metric suitable for constructing the semiclassical
wavefunction in the classically forbidden region. Thus the
Hartle-Hawking initial condition does not produce a semiclassical
wavefunction corresponding to Lorentzian $RP^3$ de Sitter.

What is the physical significance of this result? First, it indicates
that in semiclassical approximation to the Hartle-Hawking wavefunction,
the geometry of the universe is limited to a very special class of
3-manifolds that admit positive scalar curvature: ones that  divide
Einstein manifolds. What is disturbing is that, as explicitly seen for
$RP^3$, the obstruction to having a classical Euclidean path with
$RP^3$ boundary occurs at one point. This suggests that there are
manifolds that are approximately Einstein  occurring in the Euclidean
integral; that is paths with metric of form \(idacl) up to a $RP^3$
boundary of radius $a_0$ at $\tau_0$ that then become nonextremal paths
on some manifold $G$ that smoothly caps off this boundary.  Note that
the diameter of this boundary $RP^3$ can be made arbitrarily small by
taking $a_0 \to 0$.  Also, the volume of $G$ can be made arbitrarily
small by a suitable conformal transformation. Thus if the curvature of
the metric on $G$ can be controlled such that its integral over $G$ can
be made small then the action would approach an extremum. Therefore, it
appears that there are sequences of manifolds histories that approach a
limiting extremum; however, this limit point is not contained in the
space of histories itself. Thus from a physical point of view, the
contribution from the stationary point that is supposed to approximate
the contribution from these paths is being suppressed because of a
mathematical technicality; the limit point is not in the space of
histories. Therefore this semiclassical result suggests that the
properties of semiclassical histories considered in the Euclidean
functional integral should be generalized.  Furthermore, it immediately
follows that the generalization should be carried over to the space of
histories for the Euclidean functional integral itself.

{}From the topological nature of the example, it is logical to consider
generalizing the topology of the histories to be included in the
Euclidean integral. However, it is useful to first discuss the
 most immediate generalization;  to allow some set of complex metrics,
that is metrics which are neither Lorentzian or Euclidean as paths in
the Euclidean functional integral.\refto{jimhall}
 This proposal is appealing as it is intimately connected to unresolved
issues such as conformal rotation\refto{ghp} needed to concretely
implement the Euclidean functional integral beyond the semiclassical
approximation. Although the study of this approach is certainly
worthwhile, it must be defined and implemented.  A complex metric can
be considered to be a complex symmetric tensor defined on a smooth
manifold. A general complex symmetric tensor is not invertible and will
not have well defined curvature, so one immediately expects that the
generalization must be restricted to some smaller set of complex
metrics for which the action and curvature can be made well defined.
How to do so is not straightforward; the Einstein equations for a
general complex metric are neither elliptic nor hyperbolic and thus
standard existence theorems for solutions to initial data do not
apply.  Such an existence theorem cannot be constructed for general
complex metrics and thus a specification of the allowed complex metrics
must be made and an existence theorem proven for this set of complex
metrics. Such a specification can be made well defined in simple
models,\refto{jimhall} however, it remains to be seen how to do so for
some general family of complex metrics.

The immediate effect of such a generalization is that the constraint on
the scalar curvature of the boundary 3-manifold is slightly relaxed as
the continuity condition on the wavefunction \(match) no longer
involves purely real and imaginary actions. Thus, one might hope to
produce semiclassical amplitudes for a broader class of 3-manifolds.
However it is important to note that
though it may provide more general semiclassical wavefunctions, it
does not provide a method of including stationary limit points for all
sequences of histories. In particular it does not provide such a
stationary limit point for the $RP^3$ boundary with round metric. For
example, allowing for complex solutions of the form \(idacl) does not
produce a
manifold in the $RP^3$ case. Take the metric to be a one parameter
complex metric by allowing the variable $\tau$ in \(idacl) to be
complex. A direct computation of the curvature leads to the same
equations of motion \(eom) as for the real $\tau$ case. The solution to
the equations of motion is analytic; consequently, the topological
space is always pinched off at a non-manifold point. Again, as in the
real case, more general arguments can be made to show that the
non-manifold point persists for complex metrics.  Therefore complex
metrics do not provide a generic solution to the
problem; indeed this is to be expected as the topology and metric of a
history are specified independently. Thus one is brought back to the
idea that the topology of the histories should be generalized.

A natural starting point is to generalize the topology in the minimal
way needed to produce an amplitude for $RP^3$.  For example,
semiclassical calculations of the Hartle-Hawking wavefunction for a
given boundary 3-manifold could be performed on the covering space of
the boundary manifold, with the result being pulled back to the
original space. This proposal abandons the manifold restriction on the
classical Euclidean solution, replacing it with a restrictive
generalization. It immediately yields semiclassical amplitudes for
round $RP^3$; the covering space of $RP^3$ is simply $S^3$ which does
have a semiclassical amplitude. However, it is not true in general that
the covering space of an arbitrary compact 3-manifold is also a compact
3-manifold. For example, let $\Sigma^3$ be a $K(\pi,1)$ manifold, that
is a manifold whose only nonvanishing homotopy group is  the
fundamental group. The covering space of this manifold is an open
contractible 3-manifold which is not compact; therefore its covering
space is not the boundary of a compact 4-manifold.  Therefore, the
Hartle Hawking wavefunction is not defined for the covering space and
thus the calculation cannot be performed. Of course one might argue
that this case is not interesting because in general $\Sigma^3$ will
have negative $^3R$ and thus cannot be the boundary of an Einstein
manifold with $K_{ij}=0$. However, it is unsatisfactory to have a
prescription that must be applied on an ad hoc basis.  Therefore, an
appropriate generalization of the topology of the history should be
formulated in more generic terms. Certain results on the moduli space
of Einstein metrics on manifolds support such a
viewpoint.\refto{anderson} Under certain conditions on the volume and
curvature of the manifold, it can be proven  that nonmanifold Einstein
spaces occur as boundaries of the moduli space of Einstein metrics on
manifolds. Such spaces have points locally related to covering spaces;
however, globally their structure is different. Therefore,
heuristically, one expects that such spaces must be included in the
semiclassical approximation to the sum over histories. In order to
begin the definition of an appropriate set of topological spaces to
allow as histories, it is useful to summarize the properties needed for
use in Euclidean functional integrals for gravity.

\head{4.~Requirements on the Class of Generalized Histories}
\taghead{4.}

Given the motivation for including histories formed of more general
topological spaces in Euclidean functional integrals, the next step is
to address the issue of what set of topological spaces should be
included.  In order to do so, the purpose must be kept in mind, namely
formulating Euclidean functional integrals for Einstein gravity. Recall
that
 Riemannian histories played a critical role in the formulation of the
Euclidean functional integral \(hhgstate); indeed these
 classical histories carry the topological properties of the space of
 histories.  It is clear that a useful starting point for finding a
more general space of histories is to find a description of the
generalized histories that are the analogs of the Riemannian histories;
that is to find well behaved  classical histories formulated on more
general topological spaces. It is  obvious that these more generalized
histories must contain all manifolds.  The requirement that the
generalized topological spaces have  a notion of classical geometry
places certain restrictions on the kinds of topological spaces allowed.
Thus this section will discuss the geometrical properties that are
required of candidates for generalized histories; histories and their
weighting must be implementable.

A generalized history in the Euclidean gravity consists of a
topological space $X$.  This history is weighted by an action, which in
the case $X$ is a smooth manifold, is simply the Euclidean action
$I(g)$. In order to concretely define the corresponding action for the
generalized topological space $X$, distance, volume, and scalar
curvature must be defined. This restriction limits the set of
generalized topological spaces.

In order to define distance, the topological spaces $X$ must be
metrizable. \refto{spanier,rs}
 This alone is not enough of a restriction because metrizable
topological spaces can have regions of different dimension. For
example, a sphere with a flagpole attached at the north pole is a
rather nice metrizable space. However, how to weight the contributions
of such spaces in a sum over histories is not clear as the form and
properties of the action depend on the dimension of the space.
Therefore, it is reasonable to require that the dimension of the
metrizable topological space be well defined and more specifically,
that it have uniform dimension. When the dimension $n$ is well defined,
the space will be denoted by $X^n$.
 Again, one can find examples which satisfy this new condition but
which are unsatisfactory for other reasons.  So before proceeding with
the selection of the generalized space it will be useful to review some
facts about manifolds. Furthermore, by stating some of these properties
in a more abstract language, one is able to decide how to define things
such as distance, geodesics, integration, and curvature on the
generalized spaces. Once armed with these concepts the choice of
suitable topological spaces is simplified.

The length of a smooth curve on a Riemannian manifold is easily defined
using the metric and tangent vectors to the curve. Since the metric is
positive definite, the length can be used to define a distance function
on the manifold in the following way:\refto{kn}

\proclaim Definition {(4.1)}. Given a connected Riemannian manifold
$M^n$, a
distance $d(x,y)$ for $x,y\ \epsilon\ M^n$ can be defined by $$d(x,y)
= \inf_{c\ \epsilon \Omega[x,y]} \ell(c)\eqno(distance)$$ where
$\Omega[x,y]$ is the set of all $C^1$ curves between the points $x$ and
$y$, $$\Omega[x,y]= \{c:[0,1]\rightarrow M^n|c\ \ \hbox{is}
\ \ C^1\ \ \hbox{with}\ \ c(0) = x\ \ \hbox{and}\ \ c(1)= y\}$$ and
$\ell(c)$ is the length of the curve, $$\ell(c) =\int_0^1
{\bigl(g_{ab}\dot c^a \dot c^b\bigr)}^{1\over 2} ds.
\eqno(length)$$\par

\noindent The distance function $d$ defined above is compatible with
the topology of the manifold.  This means that open sets and all other
topological properties are determined entirely by $d$. Of course, the
choice of different Riemannian metrics yields different distance
functions; however all the topological properties of the manifold will
remain equivalent for all choices of metric.  Furthermore, the critical
points of $\ell$ are geodesics. This suggests a relationship between
geodesics and topology which is explicitly given in the following well
known theorem.

\proclaim Theorem (Hopf-Rinow) {(4.2)}. Given a connected Riemannian
manifold $M^n$, the following are equivalent:  \item{i)} The distance
function d(x,y) is Cauchy complete.  \item{ii)} M is geodesically
complete.  \item{iii)} Any two points can be joined by a minimizing
geodesic.\par

\noindent One can see that this theorem is useful for deciding when a
Riemannian manifold is geodesically complete. For example, given a
closed Riemannian manifold one can easily show it is complete as a
metric space using a compactness argument. Then direct application of
the above theorem proves it is geodesically complete.

One might wonder how much of the geodesic properties are described by
$d$.  In fact, geodesics can be characterized entirely in terms of $d$.
To see this, one can prove that the length of a geodesic is the
distance between its endpoints. Furthermore, given any point of the
geodesic between the endpoints,
the length of the geodesic is the sum of the distances from each
endpoint to the third point. This property of the addition of distances
is captured in the more concise definition:

\proclaim Definition {(4.3)}. Given a topological metric space $X$ with
distance function $d$, a segment is a continuous map
$c:[a,b]\rightarrow X$ such that
$$d\left(c\left(t_1\right),c\left(t_2\right)\right)
 +d\left(c\left(t_2\right),c\left(t_3\right)\right)
=d\left(c\left(t_1\right),c\left(t_3\right)\right)$$ whenever $a\le
t_1\le t_2 \le t_3 \le b$.\par

\noindent  Although not relevant to this discussion, note that
Def.(4.3) applies to metric spaces of nonuniform dimension.

A geodesic is therefore a segment. The following theorem proves the
converse; it is proven using the fact that every point has neighborhood
such that any two points in the neighborhood are connected via a
geodesic.  The interesting feature of the following theorem is that a
segment is only assumed to be continuous by Def.(4.3).

\proclaim Theorem {(4.4)}. Let $M^n$ be a Riemannian manifold with
Riemannian metric $g$ and induced topological metric $d$, the segments
of $d$ are geodesics (up to parameterization) of $g$.\par

The notion of equivalence of Riemannian manifolds is useful both
mathematically and physically. Recall, two Riemannian manifolds are
equivalent if there is a diffeomorphism between them such that the
metrics are pullbacks of one another. In order to address this issue,
one can see what happens to $d$ via the pullback. Locally, this is just
a change of coordinates and the two distances will be related via an
isometry as defined below.

\proclaim Definition {(4.5)}. An isometry between metric spaces $X$ and
$X'$ is a map $f: X\rightarrow X'$ such that $d\bigl(x,y\bigr) =
d'\bigl(f(x),f(y)\bigr)$.\par

\noindent Again, note that Def.(4.5) holds for metric spaces of
nonuniform dimension. Conversely, given any isometry of the two
distance functions as defined in \(distance), the following is true:

\proclaim Theorem {(4.6)}. Given a map $f$ of a Riemannian manifold
$M^n$ onto a Riemannian manifold $M'^n$, such that  $d\bigl(x,y\bigr) =
d'\bigl(f(x),f(y)\bigr)$, then $f$ is a diffeomorphism and $g =
f^*g'$.  Furthermore, one does not need to assume continuity of
$f$.\par

The above definitions and theorems all reflect properties of Riemannian
manifolds that are utilized in constructing the geometry of histories
for Euclidean functional integrals. Any candidate for a set of
generalized histories will also need to satisfy these properties in
order for the geometry of a history to be well behaved. Thus at this
point the set of topological spaces can be greatly restricted by
imposing the  requirement that the above theorems and definitions or a
direct generalization of them hold. Such a set of topological spaces
are polyhedra or spaces homeomorphic to a polyhedra.\refto{rs}

In order to define these spaces, a little background is needed.  The
notion of the join of topological spaces was given in Def.(3.2);
however, this definition  is only continuous and does  not impose any
further conditions. It useful to define a more restricted version of
joins in order to obtain a nicer set of spaces. This definition is the
{\it piecewise linear} or {\it PL} version. Let two subspaces $X$ and
$Y$ of ${\bf R}^n$ be positioned so that for any distinct points
$x_1,x_2\in X$ and $y_1,y_2\in Y$ the line segments connecting $x_1$ to
$y_1$ and $x_2$ to $y_2$ do not intersect. For spaces $X$ and $Y$
positioned in such a way, the {\it PL join} is the union of all line
segments joining points of $X$ to points of $Y$. The join will be
denoted $XY$. One can show that as topological spaces, the PL join when
it exists is homeomorphic to the topological join $X*Y$. This means
that as topological spaces, there is no difference between the PL join
and the topological join. The difference
 is that the PL join has more structure and can only be defined when
the two space can be positioned in this nice way.  Using the PL join, a
PL cone is defined to be PL join of space $X$ with a point $a$. It will
be denoted by $aX$. Again a PL cone $aX$ is homeomorphic to a
topological cone $C(X)$.

\proclaim Definition {(4.7)}. A polyhedron $P$ is a subset of ${\bf
R}^n$ such that each point $p$ has a cone neighborhood $N=aL\subseteq
P$ where $L$ is compact topological space. $L$ is called the link of
the neighborhood $N$.\par

\noindent It is important to note that the dimension of $P$ is not
necessarily that of the space ${\bf R}^n$. Like a smooth manifold, the
space $P$ can be triangulated. This is connected to the fact that the
spaces $L$ are not pathological and that the PL join is linear.  The
spaces $P$ given by Def.(4.7) may seem limited at first; however, any
topological space which has a triangulation is always homeomorphic to a
polyhedron. Therefore the spaces $P$ are quite general. Since this
paper
is concerned with topological properties of spaces alone, it is useful
at the present time to introduce the convention that the term
polyhedron refers to any space homeomorphic to a polyhedron.

There are many examples of polyhedra: Smooth manifolds are a subset of
polyhedra; in this case each point has a cone neighborhood where
$L\approx S^{n-1}$.  An example of a non-manifold polyhedron is the
surface of a n-dimensional hypercube with a line segment attached at
one of its vertices; in this case the vertex where the line segment is
attached has a cone neighborhood where $L$ is the disjoint union of
$S^{n-1}$ and a point,  all other points in the hypercube have cone
neighborhoods with $L\approx S^{n-1}$ and all other points in the line
segment have cone neighborhoods with $X\approx S^0$.  It is clear that
all $L$ are compact topological spaces even though they are not all
connected manifolds.

It is obvious from these examples that polyhedra include metrizable
spaces of different dimension; however it is easy to restrict to the
subset of pure polyhedra: A pure polyhedron is one for which every
point has a cone neighborhood $aL$ with the same dimension n. Other
well defined subsets of the set of all polyhedra can be formed as well
by applying the usual definitions of closed and connected to these
topological spaces.\refto{diffgeom}

Next the requirement that the action be defined is now  more
appropriately addressed as it can  be restricted to the case of pure
polyhedra.  As pure polyhedra are subsets of ${\bf R}^n$,  Lebesgue
integration of integrands with the appropriate behavior is well
defined, namely the Lebesgue integral is well defined if the integrand
is defined except on sets of measure zero. Secondly, if the notion of a
stationary point of the action is to carry over to generalized
histories, it must be possible to associate a well defined scalar
curvature with each point of the polyhedra. Thus the task is to find a
subset of pure polyhedra for which the scalar curvature has the
appropriate behavior.

Recall that the Riemann curvature of a Riemannian manifold is defined
in terms of parallel transport around infinitesimal closed curves on
the manifold. Such curves lie in a neighborhood diffeomorphic to ${\bf
R}^n$.  Given the Riemann curvature, the scalar curvature is then
computed by taking the appropriate contraction of the indices with the
metric. Thus, turning to the case of pure polyhedra, it is clear that
scalar curvature can be defined exactly as for manifolds for all points
in the polyhedra that have neighborhoods diffeomorphic to ${\bf R}^n$.
Thus the task is to extend this definition of curvature to points of
the polyhedra which do not have such neighborhoods. Note that one only
needs to extend the scalar curvature to all points in the pure
polyhedra in order that the action of the history be well defined. As
the scalar curvature must be well defined at all points in the pure
polyhedra, the set of measure zero must not have Hausdorff dimension
greater than zero; i.e. it must not be a line segment, triangle, or
other (n-1)-dimensional set. The reason is that for such sets of
measure zero, the value obtained from taking the limit of the scalar
curvature onto the set will be direction dependent for arbitrary
metrics. For example, consider a polyhedra consisting of two 2-spheres
with a line segment in each identified as in Figure 4. The space
consisting of the polyhedra minus the line segment is two disjoint
2-spheres, each minus a line segment. The scalar curvature on each two
sphere is well defined, even though the metric on the space is
incomplete. However, the scalar curvature does not approach a well
defined value as points on the missing line segment are approached for
a general metric on the space. The value of the scalar curvature will
depend both on the direction of approach and on the parameterization of
the removed line segment. Thus a limiting procedure  will not yield a
well defined scalar curvature on each point of the line segment.
Therefore  the subset of pure polyhedra must be restricted to a subset
for which there is a set of isolated points whose cone neighborhoods
$aL$ are not homeomorphic to ${\bf R}^n$; additionally, the closed
spaces $L$ must be manifolds. In this case, the limit of the scalar
curvature onto the singular points is controlled by the relation of the
metric on the space to the distance function; away from the singular
point $a$, the curvature is continuous as the cone neighborhood $aL$
minus the point $a$ is a manifold and as the point $a$ is approached,
the curvature must approach a fixed value from all directions.

Even though this subset of pure polyhedra forms a relatively nice set
of topological spaces, the scalar curvature still cannot necessarily be
defined at all points for all polyhedra in this subset. For example,
consider the compact polyhedron formed by taking two $C(S^3)$ and
identifying the two vertices as illustrated in Figure 4.   Let the
metric on the first cone be $dt^2 + \alpha^2t^2 d\Omega^2 $ and that on
the second be $d{\rho }^2 + \beta^2{\rho }^2d\Omega^2 $ with constants
$\alpha \neq \beta$ and the vertex at $t={\rho }=0$. The scalar
curvature can be computed on each cone away from the vertex; it is
$6/\alpha^2$ on the first and $6/\beta^2$ on the second. However, as
the scalar curvature is not the same constant on both cones, it cannot
be defined at the vertex by an extension of the function on both cones
to this point. Therefore, the scalar curvature is not necessarily well
defined at the vertex.  This example indicates that the subset of pure
polyhedra should be further restricted to be those for which the
isolated set of points have cone neighborhoods $aL$  where $L$ is a
closed connected manifold.

Thus, geometrical properties greatly restrict the set of topological
spaces that can be considered as candidates for  generalized histories.
It is now possible to propose and discuss a new set of generalized
histories that is a subset of these polyhedra, contains all manifolds
and contains the suspension of $RP^3$ that appeared in the study of the
semiclassical approximation.
The geometry of the above set of polyhedra will be done by using the
properties which only depend on the distance function.  In particular,
given a Riemannian metric at the manifold points of the polyhedra one
will choose a distance function on the polyhedra which is the Cauchy
completion of the distance induced by the given Riemannian metric. The
allows natural extensions of geometry to the new set of spaces. For
example, geodesics on these spaces will be defined as segments. By the
above theorems, this will agree with the usual definition of geodesics
at the manifold points. Similarly, these polyhedra will be considered
to be geometrically equivalent when they are isometric via an isometry
of the distance function. Again, at manifold points this is equivalent
to diffeomorphism equivalence of Riemannian metrics as mentioned above.
Integration of functions on this set of spaces will be defined via the
measure constructed from the distance function. If one deletes all of
the nonmanifold points, this is equivalent to the integration of
functions using the volume element of the given Riemannian metric.
Hence, one can define square integrable functions and other useful
objects from analysis on this set of polyhedra. Thus the study of the
geometry of spaces in this new set of spaces is akin to that of
manifolds.

\head{5.~Generalized Topological Spaces for Quantum Gravity}
\taghead{5.}

This section will present a new set of topological spaces for use as
generalized histories. This set has not been previously defined or
studied in the literature. These spaces will be called conifolds.
Smooth conifolds are a subset of the special set of pure polyhedra for
which the neighborhood of every point is a  homeomorphic to a PL cone
over a closed connected manifold.  By definition this subset clearly
includes all manifolds. As for manifolds, it is the subset of
topological conifolds that admit a smooth structure, smooth conifolds,
that are appropriate for physics.

 At this point it is useful to define a slightly more general set of
spaces than the above polyhedra by weakening the PL cone neighborhoods
to be only the topological cones $C(L)$. This generalization allows the
following definitions to parallel the analogous definitions for
manifolds. The name of this more general set, conifolds, is due to the
observation that the neighborhood of every point is a cone.
\proclaim Definition {(5.1)}. A n-dimensional conifold $X^n$, $n\ge 2$,
is a metrizable space such that given any $x_0 \in  X^n$ there is an
open neighborhood $N_{x_0}$ and some closed connected (n-1)-manifold
$\Sigma_{x_0}^{n-1}$ such that $N_{x_0}$ is homeomorphic to the
interior of $C(\Sigma_{x_0}^{n-1})$ with $x_0$ mapped to the apex of
the cone. Any neighborhood homeomorphic to such cones will be referred
to as conical neighborhoods.

A zero dimensional conifold will defined as a collection of discrete
points. This is the same definition as that of a  zero dimensional
manifold.  The definition of a one dimensional conifold is almost the
same as that of
the above n-dimensional conifold except that the links are assumed to
have two disconnected components.  The reason for this  follows from
the case of 1-manifolds: Every point of a 1-manifold has a neighborhood
that is a cone over the zero sphere $S^0$. Since $S^0$ consists of two
points it has two disconnected components; consequently the
neighborhood of any point of a 1-manifold is a cone over two
disconnected points. Thus it is natural to define a 1-conifold by
requiring that the links consist of  two disconnected components.
Using this definition, 1-conifolds are just 1-manifolds.
Therefore, zero and one dimensional conifolds are manifolds.

In two dimensions, the general n-dimensional definition applies,
however all 2-conifolds are manifolds because the only closed connected
1-manifold is the circle $S^1$. Since there are a countably infinite
number of closed connected 2-manifolds, 3-conifolds are not just
3-manifolds but more general metrizable spaces. The same will be true
in all dimensions larger than three.  The set of all n-conifolds
includes all topological n-manifolds as the neighborhood of every point
in a manifold is homeomorphic to a cone over $S^{n-1}$. Secondly it
follows from Def.(3.4) that the set of n-conifolds includes the
suspensions of all closed connected (n-1)-manifolds. For example, the
suspension of $RP^3$ described in section 3 and illustrated in Figure 3
is also included in the set of 4-conifolds; the neighborhoods of the
singular points at the north and south poles are cones over $RP^3$ by
construction and all other points have neighborhoods homeomorphic to
cones over $S^3$. This space is an example of a conifold with a finite
number of nonmanifold points.  Closed n-conifolds which are neither
manifolds nor suspensions of manifolds can also be readily constructed.
Two examples are given below and Figure 5 provides a visualization of
another.

The first example is $K^n = T^n/{\bf Z}_2$ for $n\ge 3$ where the ${\bf
Z}_2$ action is defined as follows: The n-torus  can be thought of as
$T^n = \{(\hbox{\rm e}^{i\theta_1},\hbox{\rm
e}^{i\theta_2},\ldots,\hbox{\rm e}^{i\theta_n}) |\  -\pi\le \theta_n<
\pi\}$. Define a self-homeomorphism $c:T^n \to T^n$ by $c(\hbox{\rm
e}^{i\theta_1},\hbox{\rm e}^{i\theta_2},\ldots,\hbox{\rm
e}^{i\theta_n})= (\hbox{\rm e}^{-i\theta_1},\hbox{\rm
e}^{-i\theta_2},\ldots,\hbox{\rm e}^{-i\theta_n})$.  Clearly, $c$ is a
self-homeomorphism and $c^2$ is the identity map. Hence $c$ gives an
action of the cyclic group ${\bf Z}_2$ on $T^n$. Finally $K^n$ is
defined as the quotient space $T^n/{\bf Z}_2$ where ${\bf Z}_2$ is
represented by $c$.

$K^n$ as just defined is a closed conifold with $2^n$ singular points
for $n\ge 3$ and each singular point has a conical neighborhood
homeomorphic to a cone over $RP^{n-1}$. In order to prove this, the
observation that the identifying map $c$ has fixed points is utilized.
In general, such fixed points indicate that $K^n$ will not be a
manifold. Let $F= \{x\in T^n | \ c(x)=x\}$, i.e. the set of fixed
points of the ${\bf Z}_2$ action. Next pick some $x_0 \in F$ and a
small closed neighborhood $B^n_{x_0}$  of $x_0$ in $T^n$ such that
$B^n_{x_0} \cap F = \{x_0\}$ and $B^n_{x_0}$ is homeomorphic to a
n-ball with a (n-1)-sphere as boundary. Next, define a new neighborhood
$N^n_{x_0}$ of $x_0$ by $B^n_{x_0} \cap c(B^n_{x_0})$ where
$c(B^n_{x_0})$ is the image of $B^n_{x_0}$ with respect to $c$. By
construction, $c(N^n_{x_0}) = N^n_{x_0}$ and $N^n_{x_0}$ is
homeomorphic to a n-ball.  Now let $Y^n= T^n-\bigcup_{x\in F} {\rm
int}(N^n_x)$. The ${\bf Z}_2$ group action acts freely on $Y^n$ because
all fixed points are removed. Hence $Y^n/{\bf Z}_2$ is a compact
manifold with boundary. Since the boundary of $Y_n$ is a disjoint union
of (n-1)-spheres and ${\bf Z}_2$ acts freely on each, it follows that
$Y^n/{\bf Z}_2$ is a compact manifold with boundary equal to the
disjoint union of $2^n$ copies of $RP^{n-1} = S^{n-1}/{\bf Z}_2$.
Finally, $K^n = Y^n/{\bf Z}_2\cup N(S)$ where $N(S) = \bigcup_{x\in F}
N^n_x/{\bf Z}_2$. Each  $N^n_x/{\bf Z}_2$ is homeomorphic to ${\bf
Z}_2$ acting on a n-ball with one fixed point and it follows that
$N^n_x/{\bf Z}_2$ is a cone over $RP^{n-1}$. Consequently $K^n$ is a
conifold for $n>2$; $ K^2$ is a manifold because $RP^1 = S^1$.

The second example is $L^{2n} = CP^n/{\bf Z}_3$ for $n\ge 2$ where
$CP^n$ is complex projective space and the ${\bf Z}_3$ action will be
defined below. $L^{2n}$ is a closed conifold with one singular point.
$CP^n$ is the set of complex lines in ${\bf C}^{n+1}$; however, a more
useful description of it for present purposes is the following: Let the
(2n+1)-sphere $S^{2n+1}$ be defined by $S^{2n+1} =\{ (z_0,z_1,\ldots,
z_n) |\ z_k \in {\bf C}^{n+1}, \sum_{k=0}^n {\bar z_k} z_k =1\}$. Now
define the equivalence relation $(z_0,z_1,\ldots, z_n)\sim
(w_0,w_1,\ldots, w_n)$ if and only if $(z_0,z_1,\ldots, z_n)=
(\hbox{\rm e}^{i\theta} w_0,\hbox{\rm e}^{i\theta}w_1,\ldots, \hbox{\rm
e}^{i\theta}w_n)$ for some real $\theta$.  Then $CP^n =
S^{2n+1}/\sim$.

Next let $a: CP^n \to CP^n$ be defined by $a([z_0,z_1,\ldots, z_n]) =
[az_0, z_1,\ldots,z_n]$ where $a = \hbox{\rm e}^{\frac {2\pi i}3}$ and
$[z_0,z_1,\ldots, z_n]$ is the equivalence class in $S^{2n+1}/\sim$.
Clearly, $a$ is a self-homeomorphism and $a^3$ is the identity map.
Using the above definitions combined with some basic algebra, it is
possible to show that the only fixed point is the point
$[1,0,\ldots,0]$ in $CP^n$.  Repeating the same type of analysis as
done in the case of $K^n$, one can show that $L^{2n}$ is a conifold
with one singular point with neighborhood $S^{2n-1}/{\bf Z}_3$.

Examples of n-conifolds with an infinite number of nonmanifold points
can also be constructed: Pick a countably infinite number of disjoint
n-balls in ${\bf R}^n$ with $n\ge 3$, and continuously embed a
(n-1)-torus in each ball. Now, remove the interior region bound by each
torus. The resulting space is a n-manifold with a countable number of
(n-1)-tori as boundary. Finally, cap off the boundaries on this
manifold with the cone over each torus.  The resulting space is a
n-conifold.

In the above examples, the conifolds are manifolds except at a discrete
set of points. In general, this will be true for all conifolds. In
order to see this, let $S$ be the set of points in a n-conifold $X^n$
which do not have neighborhoods homeomorphic to the interior of a cone
over $S^{n-1}$.  The set $S$ is called the {\it singular set} of $X^n$.
It is the set of points at which the conifold is not a manifold. In
order to prove that $S$ only consists of a discrete set of points, one
needs to show that $S$ has no limit points, i.e. that one can find a
collection of disjoint neighborhoods around all of the points in $S$
simultaneously. Each $x_0\in S$ has a conical neighborhood $N_{x_0}$
homeomorphic to $C(\Sigma_{x_0}^{n-1})$. Since the only nonmanifold
point of the cone is the apex, it follows that the only nonmanifold
point of $N_{x_0}$ is $x_0$. Hence, as each $x_0$ in $S$ has a
neighborhood that contains no other point of $S$, no point of $S$ is a
limit point of the set $S$. Now, one must show that the neighborhoods
can be chosen to be simultaneously disjoint; this is equivalent to
showing that there are no limit points of $S$ in $X^n$.  Let $x_1\in
X^n$ be some point not in $S$. This means $x_1$ has a conical
neighborhood homeomorphic to a cone of a (n-1)-sphere. Hence all points
in this neighborhood are manifold points. Hence, as $x_1$ has a
neighborhood that contains no point in $S$, $x_1$ is not a limit point
of $S$. Therefore, $S$ has no limit points. Immediately it follows that
$S$ consists of a discrete set of points.

The set $S$ of singular points is always countable for each connected
component of a n-conifold. In order to see this, let $X^n$ be a
connected conifold.  Let $N(S)$ be the disjoint union of connected
conical neighborhoods about each of the singular points. From the above
discussion, $X^n-N(S)$ is a topological connected n-manifold with
boundary and therefore can be continuously embedded in ${\bf
R}^{2n+1}$. If one continuously pinches off each of the boundaries of
$X^n-N(S)$, one obtains a space homeomorphic to $X^n$. Hence,
n-conifolds can be continuously embedded in ${\bf R}^{2n+1}$. Now, the
topology of ${\bf R}^{2n+1}$ has a basis which consists of a countable
number of open sets with compact closure. It follows that $X^n$ has the
same property because it is a subspace. Now suppose that $S$ consisted
of an uncountable number of points; as the basis of $X^n$ is countable
it follows that there would have to be an infinite number of points of
$S$ in one of the these sets with compact closure. This would mean that
$S$ has limit points; however by the proof of the previous paragraph,
$S$ can have no limit points.   Therefore, the set $S$ can have only a
countable number of points.

Def.(5.1) defines topological n-conifolds without boundary; it is also
useful to define n-conifolds with boundary. The boundary of an
n-conifold will be required to be a (n-1)-manifold without boundary.
This restriction is consistent with the definition of n-manifolds with
boundary; additionally it
makes the mathematics of these spaces less pathological. It is also
motivated by the requirements of a physical application of conifolds to
Euclidean functional integrals as the argument of the amplitudes
typically consists of a metric  on a closed (n-1)-manifold.  Thus,
specifically, a {\it n-conifold with boundary} is a metrizable space
$X^n$ such every point has a conical neighborhood as in Def.~(5.1) or
an open neighborhood homeomorphic to open subset of the half-space
${\bf R}_+^n$. The {\it boundary} of $X^n$, $\partial X^n$, is defined
as the set of points which are mapped to boundary points of ${\bf
R}_+^n$. Again, the singular set $S$ is defined as the set of points in
$X^n$ whose conical neighborhoods are not homeomorphic to a n-ball.
The definition of a n-conifold with boundary
implies that all the  singular points are on the interior of $X^n$.
Again, one can show that on each connected component of $X^n$, $S$ will
be a countable set. Similarly, $X^n-S$ will be a manifold with boundary
and $\partial X^n=\partial (X^n-S)$. This property implies that the
boundary of a conifold is topologically well defined; it is preserved
under homeomorphisms of the space. Finally, in analogy with the
manifold case,  a {\it closed conifold} is defined as a compact
conifold without boundary.

Given that three dimensions is the lowest dimension allowing nontrivial
examples of conifolds, it is interesting to see how different the set
of 3-conifolds is from that of 3-manifolds. Also, it would be useful to
have a simple test that describes the subset of 3-conifolds which are
3-manifolds. Recall that the Euler characteristic of a topological
space is just the alternating sum of its Betti
numbers.\refto{footnote1}
Thus calculation of the Euler characteristic  is
reduced to simple arithmetic for such spaces. The observation that the
Euler characteristic is zero for closed 3-manifolds suggests that it
provides a simple test for determining whether or not a 3-conifold is a
3-manifold.  Indeed, the following theorem will be of great value in
Part II of this paper to discuss the algorithmic decidability of
conifolds.

\proclaim Theorem {(5.2)}. Let $X^3$ be a closed 3-conifold. Then $X^3
$ is a 3-manifold iff $\chi(X^3) = 0.$\par

\noindent  First $X^3$ is a  3-manifold except at a finite set of
singular points $S$.  At each singular point in $S$, choose a small
connected neighborhood such that it is homeomorphic to a cone over some
closed 2-manifold. Let the collection of all these neighborhoods be
$N(S)$.  Then $M_0\equiv X^3 - N(S)$ is a compact 3-manifold with
boundary.  Using the Mayer-Vietoris sequence for homology one can show
that $$\chi(X^3) -\chi(M_0) + \chi(\partial M_0) - b_0(S) =
0\eqno(euler1)$$ and also as demonstrated in Appendix  A that $$
2\chi(M_0) = \chi(\partial M_0).\eqno(euler2)$$ Combining these two
equations yields $\chi(X^3) + \12 \chi(\partial M_0) = b_0(S).$
 Now, assume that $\chi(X^3) = 0$; then $$\chi(\partial M_0) =
2b_0(S).\eqno(boundaryeuler)$$ Next $b_0(\partial M_0) = b_0(S)$ as
$b_0(\partial M_0)$ is the number of connected components of  $\partial
M_0$. Furthermore, $b_2(\partial M_0) \le b_0(S)$ because on each
connected component of $\partial M_0$, $b_2$ is either one or zero.
Hence, using these two results and $\chi(\partial M_0) = b_0(\partial
M_0) - b_1(\partial M_0) + b_2(\partial M_0)$ in \(boundaryeuler),
$$-b_1(\partial M_0) = b_0(S) - b_2(\partial M_0) \ge 0.$$ The only
solution is if $b_1(\partial M_0) = 0$. This implies that $b_2(\partial
M_0)= b_0(S)$. Finally, it follows that the Euler characteristic of
each connected component of $\partial M_0$ is equal to $2$. The only
closed connected 2-manifold with Euler characteristic equal to 2 is a
2-sphere.  Hence $N(S)$ is the disjoint union of 3-balls. Therefore,
$X^3$ is a 3-manifold.

\noindent Conversely, it is easily proven that the Euler characteristic
of any odd-dimensional manifold is zero. Q.E.D.

The proof of this theorem can be applied directly to compute the Euler
characteristic for explicit constructions of 3-conifolds; for example
one can show from the construction of $K^3=T^3/{\bf Z}_2$ that it is a
3-conifold.  Recall that $K^3 = Y^3/{\bf Z}_2 \cup N(S)$ where $S$ is
the set of 8 singular points and $Y^3/{\bf Z}_2$ is a compact manifold
with boundary consisting of the disjoint union of eight $RP^2$
manifolds. Eqn.\(euler2) can be applied to find that $2\chi(Y^3/{\bf
Z}_2) = 8\chi(RP^2) $. Then,  using $\chi(RP^2) =1$, \(euler1)
immediately yields $\chi(K^3) = -4\chi(RP^2)  + b_0(S) = 4$. Thus the
proof of Thm.(5.2) provides a very useful method of computing the Euler
characteristic  for certain 3-conifolds as well as a characterization
of the subset of 3-manifolds.

In four dimensions, conifolds are more complicated than in three
dimensions. For example, there is no simple test to determine which
4-conifolds are 4-manifolds as in Thm.(5.2) as there is no known
algorithmic description of 4-manifolds.\refto{IIa} Another problem in
four or more dimensions is that there exist conifolds which are not
homeomorphic to  polyhedra. This may seem strange until one recalls
that there are closed topological 4-manifolds which are not
homeomorphic to polyhedra. Such examples will be discussed more fully
in Part II of this paper after the appropriate machinery is introduced;
however, the point is that examples of poorly behaved conifolds exist
in four or more dimensions and this means that one  needs to place
extra conditions on topological conifolds in order to use them in
physics. This is not a conceptual problem as such extra conditions  are
indeed imposed in the case of manifolds already as discussed in section
2;  some sort of smoothness or regularity is required if
 geodesics, triangulations, and other geometric objects are to be
definable. Thus such extra conditions are natural. Therefore given the
topological definition of conifolds, one would like to discuss the
smoothness and geometry of conifolds. Again as in the case of
manifolds,
conifolds with boundary and conifolds without boundary can be
discussed at the same time.

One can define an atlas on a conifold in exactly the same manner as one
defines an atlas on a manifold:

\proclaim Definition {(5.3)}. An atlas on a n-conifold $X^n$ is a
collection $\{(U_\alpha,\varphi_\alpha)\}_{\alpha\epsilon\Lambda}$ of
open sets and homeomorphisms indexed by a set $\Lambda$ satisfying the
following:  \item{1.} The sets $U_\alpha$ cover $X^n$.  \item{2.}
$X^n-S=\hbox{\lower.9ex\hbox{${\bigcup \ \ }\atop{\alpha\in
\Lambda_0}$}}U_\alpha$ for some subset ${\Lambda}_0\subset{\Lambda}$.
\item{3.} For $\alpha\in {\Lambda}_0$, $\varphi_\alpha$ is a
homeomorphism of $U_\alpha$ to an open set in ${\bf R}^n_+$.  \item{4.}
For each $\alpha\in \Lambda -{\Lambda}_0$, $U_\alpha$ is a conical
neighborhood of a singular point and $\varphi_\alpha$ is homeomorphism
onto the interior of a cone.\par

\noindent Using the above notion of atlas, one can define a smooth
conifold by analogy with the definition of a smooth manifold Def.(2.1).
The main difference lies in defining the smoothness near the singular
points. This is done by requiring that there is a neighborhood of each
singular point such that removing it yields a smooth manifold with
boundary. More precisely,

\proclaim Definition {(5.4)}. A n-conifold is smooth ($C^k$) if and
only if there is an atlas
$\{(U_\alpha,\varphi_\alpha)\}_{\alpha\epsilon\Lambda}$ such that
$$\varphi_\beta\varphi^{-1}_\alpha:\varphi_\alpha(U_\alpha\cap
U_\beta)\rightarrow\varphi_\beta(U_\alpha\cap U_\beta)$$ is a smooth
($C^k$) map on ${\bf R}^n_+$ for $\alpha\in \Lambda_0$ and the
remaining sets $U_{\alpha}$ are connected conical neighborhoods of the
singular points such that
$X^n-\hbox{\lower.9ex\hbox{${\bigcup\ \ \ }\atop{\alpha \in {\Lambda -
\Lambda}_0}$}}U_\alpha $ is a smooth ($C^k$) submanifold of $X^n - S$
with respect to the differential structure given by
$\{(U_\alpha,\varphi_\alpha)\}_{\alpha\epsilon\Lambda}$.\par

\noindent It is important to note that an atlas defining a smooth
conifold has a subset of neighborhoods that are open sets in ${\bf
R}^n_+$ as noted in its definition.  The reason for this condition is
that smoothness for manifolds is defined in terms of the images of the
overlap of the open sets on ${\bf R}^n_+$. Thus Def.(5.4) is a close
parallel of the definition of smoothness for manifolds.

One might try to define an atlas as a cover of a conifold by conical
neighborhoods, however such an approach encounters technical problems
avoided by Def.(5.4).  For example, consider the closed conifold of
section 3 consisting of the suspension of $RP^3$. This conifold can be
covered with two open sets, each homeomorphic to a cone over $RP^3$.
This is a cover of the space; however, the intersection of the two open
sets is homeomorphic to a product of an open interval with $RP^3$
instead of to an open neighborhood in ${\bf R}^4_+$. Thus the smooth
structure cannot be defined in the usual way using this cover. In order
to define a smooth structure using these sets,  one would have provide
a new procedure to do so. For example one could  pick a smooth
structure on the product manifold given by the intersection of the sets
and then require the composition of maps on this overlap be smooth.
However, there is no natural way to do this in general as it relies on
the choice of a smooth structure on the intersection and there is no
unique or well established way of making this choice on an arbitrary
product manifold. One difficulty is
that there may be more than one smooth structure on an arbitrary
product manifold; for example, it has been shown that the product of an
open interval with a 3-sphere has an uncountable number of
nondiffeomorphic smooth structures. Another difficulty is that even if
one selects a particular smooth structure to use, one must show that
the use of this choice leads to all inequivalent smooth structures on
the conifold. Therefore, such an alternate procedure involves
unnecessary complications that can be avoided by choosing an
appropriate atlas on the conifold. Thus it is safe to say that
Def.(5.4) is a reasonable and pragmatic definition of a smooth
conifold.

For two or fewer dimensions, all conifolds are manifolds. Hence, all
n-conifolds with $n\le 2$ admit smooth structures. Similarly every
3-conifold has a smooth structure. In order to see this, remove a
conical neighborhood around each singular point of the 3-conifold, so
that the resulting space is a 3-manifold with boundary. Now, all
3-manifolds admit a smooth structure so pick one such smooth structure
on the manifold. Next  smoothly pinch off the boundary. One produces a
smooth 3-conifold this way which is homeomorphic to the original space.
Thus this generates a smooth structure on the 3-conifold. In higher
dimensions, one can construct examples of conifolds which do not admit
smooth structures.  Recalling the results for manifolds from section 2,
this should be no surprise; as there are n-manifolds for $n\ge 4$ which
admit no smooth structures, it follows immediately that there are
n-conifolds for $n\ge 4$ which admit no smooth structures. Again smooth
structures will be discussed in more detail in Part II of this paper
after the appropriate machinery is introduced.

Having defined the notion of a smooth n-conifold, its differential
geometry is done by studying the differential geometry of the manifold
$X^n-S$ and then requiring that the objects in question extend to $X^n$
in a continuous way. For example, the notion of metric on a manifold
can be extended to a metric on a conifold as below;

\proclaim Definition {(5.5)}. A Riemannian metric on a conifold $X^n$
is a Riemannian metric g on $X^n-S$ and a Cauchy complete distance
function $d$ on $X^n$ such that the distance function associated with
$g$ is equal to $d$ when restricted to $X^n-S$.\par

In other words, $d$ in the above definition is just the Cauchy
completion of the distance associated with $g$. Given the link between
geodesic completeness and Cauchy completeness of the distance on
Riemannian manifolds, it is reasonable to define the Riemannian metric
on the conifold by using a similar connection. For notational
convenience, when referring to Riemannian metrics on conifolds only the
metric on the manifold will be mentioned. It will be understood that
the distance at singular points is given by the Cauchy completion.

Having defined a Riemannian metric, defining geodesics is next.
\proclaim Definition {(5.6)}. A geodesic on a conifold is a continuous
curve which is a segment as defined in Def.(4.3).\par \noindent Recall
that segments are equivalent to geodesics in Riemannian manifolds; thus
this definition is a natural generalization of the definition of
geodesics to conifolds. As is often the case,  it is very useful to
have other equivalent characterizations of geodesics in conifolds.
Observe that if the segment never passes through a singular point of
the conifold $X^n$, then it must be a geodesic in the usual sense for
the Riemannian manifold $X^n-S$. Thus, the issue of interest is to have
an equivalent characterization of geodesics in conifolds which pass
through singular points. In order to do this, one can define a class of
curves on conifolds for which the usual length as used for curves on
manifolds makes sense.  Since the usual definition of length \(length)
involves a derivative of the curve, it is useful to use an equivalent
characterization of length in order to allow for the fact that
derivatives will not exist in a classical sense as the singular point
is approached. Before proceeding, a trick using embeddings of conifolds
in Euclidean space will be introduced. This embedding will be used to
produce an easily visualized equivalent characterization of length and
distance.

It has already been shown that all connected n-conifolds embed
continuously in ${\bf R}^{2n+1}$. One can similarly show that all
smooth n-conifolds embed smoothly in ${\bf R}^{2n+1}$. This is done
just as in the continuous case by removing neighborhoods of the
singular points and then applying the embedding theorem for smooth
manifolds, namely that  every n-manifold smoothly embeds in ${\bf
R}^{2n+1}$. Finally, the boundaries can be smoothly pinched off so that
the resulting space is equivalent to the original conifold. These
embeddings are not necessarily isometric, that is an arbitrary metric
on the conifold is not in general induced by the standard metric on
${\bf R}^{2n+1}$ by the embedding. However, a very impressive theorem
for Riemannian manifolds is the Nash embedding theorem:\refto{nash}
Given any connected Riemannian manifold $M^n$ there is some $m$ such
that $M^n$ isometrically embeds in ${\bf R}^m$. Using this theorem, one
can show that every connected n-conifold isometrically embeds in some
${\bf R}^m$; that is the conifold can be considered as a closed subset
of ${\bf R}^m$ such that the Riemannian metric on  the conifold is
induced by the usual Riemannian metric on ${\bf R}^m$. To see this,
remove all singular points $S$ in the conifold and apply the Nash
embedding theorem to the resulting manifold $X^n-S$. Now, $X^n-S$ is
isometrically embedded. However, note that the distance function as
defined by \(distance) is no longer Cauchy complete because the
singular points have been removed. Hence, by taking the completion of
the distance function, the singular points $S$ are added. Once the
singular points are included, it follows that this embedding is an
isometric embedding of the conifold $X^n$.

Since every connected Riemannian conifold can be isometrically embedded
in ${\bf R}^m$, one can assume a connected Riemannian n-conifold $X^n$
is a closed subset of ${\bf R}^m$. Now the set of {\it rectifiable
curves}, that is all  curves in ${\bf R}^m$ for which the integral
defining their length converges, can be used to define the length of
curves in $X^n$ in the following way: Given any such rectifiable curve
which is contained entirely in $X^n$, then its length is equal to its
length in ${\bf R}^m$.  This will agree with the usual length \(length)
for all curves that do not pass through singular points but more
importantly it is well defined even for those that do. Furthermore,
this is an intrinsic property of the geometry of conifolds; it does not
depend on the embedding, even though the embedding trick is a
convenient way of describing these rectifiable curves.

Moreover, as a conifold can be considered to be a closed subset of
Euclidean space and the length functional of any rectifiable curve is
bounded below, there will be a minimizing curve between any two points
which can be joined by a rectifiable curve. Thus, if one can show that
any two points in a connected Riemannian conifold can be joined by at
least one rectifiable curve, then it will immediately follow that any
two points can be joined by a minimizing curve.

Now the first step toward this result will be derived: Given any two
points in a connected Riemannian conifold $X^n$ with distance function
$d$, there is a rectifiable curve connecting them. For convenience,
assume that there is one singular point $p$; the generalization to a
countable set of singular points $S$ follows immediately.
 First observe that there is always a smooth curve connecting any two
nonsingular points because $X^n-\{p\}$ is a smooth manifold.  Next
assume that there is  no rectifiable curve passing through $p$. Then
any two points in any neighborhood of $p$ can be connected by a
geodesic in the manifold $X^n-\{p\}$; otherwise there would be a family
of smooth curves of decreasing length which limit onto a rectifiable
curve passing through $p$. This implies that $X^n -\{p\}$ is
geodesically complete; thus by the correspondence between geodesic
completeness and Cauchy completeness for smooth manifolds, $d$ must be
complete on $X^n-\{p\}$.  It follows that the singular point $p$ is not
the completion of the distance function on $X^n-S$, a contradiction to
the definition  of the conifold metric (5.5). Hence, the
 singular point $p$ must have at least one rectifiable curve passing
through it.  Finally, it follows that any two points can be connected
by a rectifiable curve. Thus, from the previous discussion, any two
points on a conifold $X^n$ can be joined by a minimizing curve.

Thus it follows from the above paragraph that
\proclaim Theorem {(5.7)}. The distance between two points in a
Riemannian conifold $X^n$ is the length of a rectifiable curve joining
them which has minimal length.\par

\noindent Furthermore, using this
result, the next theorem  follows by the same argument as in the case
of smooth manifolds and gives an equivalent characterization of
geodesics in conifolds. First note that a curve is said to be a {\it
locally minimizing} curve if the curve is minimizing for any two points
on the curve that are sufficiently close to each other. Then
\proclaim Theorem {(5.8)}. A  segment is a locally minimizing curve.\par

\noindent Let $x(t), \ t_0\le t\le t_1$ be a segment. Note that $x(t)$
is also a segment on $ t_\alpha \le t \le t_\beta $ for any $t_\alpha$
and $t_\beta$ between the endpoints $t_0$ and $t_1$. Now take
$t_\alpha$ and $t_\beta$ such that $t_\beta - t_\alpha$ is a
sufficiently small interval. By choosing this interval to be small
enough, the endpoints $x(t_\alpha)$ and $x(t_\beta)$ can be connected
with a unique minimizing curve by the previous result. Suppose for some
$t'$ between the endpoints, $t_\alpha\le t' \le t_\beta$, $ x(t')$ is
not on the  minimizing curve.
Then the distance between the two endpoints must be strictly less than
the sum of the distances from each endpoint to $x(t')$;
$$d(x(t_\alpha),x(t_\beta))< d(x(t_\alpha),x(t'))
+d(x(t'),x(t_\beta)).$$ However, this contradicts the assumption that
$x(t)$ is a segment. Therefore $x(t)$ must be a minimizing curve
between $x(t_\alpha)$ and $x(t_\beta)$. Finally note that this argument
applies to all sufficiently close points $x(t_\alpha)$ and $x(t_\beta)$
on the segment. Thus $x(t)$ is a locally minimizing curve. Q.E.D.

Conversely, a locally minimizing curve is a segment. Therefore,
geodesics in conifolds are locally minimizing curves. Also, the above
implies the following theorem:  \proclaim Theorem {(5.9)}. Any two
points of a conifold $X^n$ with a Riemannian metric can be joined not
only by a geodesic but by a minimizing geodesic.\par \noindent Observe
that this theorem is consistent with the condition that the distance
function associated with the Riemannian metric on a conifold was taken
to be Cauchy complete. Finally, it is important to stress that
 although Thms.(5.7-9) were proven using an embedding trick, they
 actually reflect intrinsic properties of the conifold. With a little
more work, these theorems can all be proven intrinsically using
 the definition of the generalized derivative on the conifold. However,
the embedding trick allows for a more intuitive understanding of these
results.

It is worth mentioning  another useful observation related to the
embedding technique; namely, one can define the tangent space at any
point of a conifold using the definition of a tangent cone for any
subset $E$ of Euclidean space:  \proclaim Definition {(5.10)}. Given
any subset $E\subseteq {\bf R}^m$, the tangent cone at a point $x_0$ is
$$\hbox{Tan}(x_0,E)=\{x\in {\bf R}^m| \ x=cr\ {\rm where}\  c\in {\bf
R}\ \ {\rm with}\ c\ge 0\ {\rm and}\ r\in {\overline T}\}$$ \noindent
where $\overline T$ is the closure of $T$ given by
$$T=\hbox{\lower.9ex\hbox{${\cap\ \ }\atop{\epsilon >0}$}} \{
{{x-x_0}\over {|{x-x_0}|}}|\ x,x_0\in E \ {\rm
and}\ 0<|{x-x_0}|<\epsilon \}.$$

\noindent At manifold points this definition yields the usual tangent
space, however, at singular points it yields a cone. Also, observe that
the tangent cone is {\it not} a vector space at singular points. One
can show that the tangent cones of conifolds have many properties in
common with the tangent spaces of manifolds; for example, even though
the tangent cone is not a vector space at the singular points, it
always has the same dimension as the conifold. Also, given nice
mappings between two conifolds, the respective tangent cones are mapped
to one another via the generalized derivative of the map.

Finally note  that spin structures and thus spinors can be defined on a
conifold $X^n$ by defining a spin structure on $X^n -S$ where $S$ is
the singular set and requiring that each $\Sigma^{n-1}_x$ for  $x\in S$
have an induced spin structure consistent with $X^n -S$. The actual
bundle of spinors on $\Sigma^{n-1}_x$ form a subspace of the bundle of
spinors on the conical neighborhood $N_x=C(\Sigma^{n-1}_x)$. Just as in
the case of manifolds, not all conifolds admit a spin structure.
Finally, for $X^n$ that do admit spin structures, once a spin structure
is chosen on the conifold, the Dirac operator can be defined by
defining it in the usual way on $X^n-S$.  These properties of conifolds
and their uses in the study of the geometry of conifolds are not used
in the present paper so details will be presented elsewhere.

\head{6.~Einstein Conifolds}
\taghead{6.}

Given the definition of the topology and geometry of conifolds it is
now possible to discuss  Einstein conifolds and their relevance to
semiclassical approximations of sums over histories for Euclidean
gravity. Recall from Def.(5.5) that the geometry of a conifold is
determined from the metric $g$ on $X^n-S$ by completion. Similarly,
other quantities such as   scalar fields and scalar curvature can be
defined on conifolds in a similar fashion: The quantities are defined
as in the case of manifolds on $X^n-S$ and their values are then
extended to the singular points $S$. A definition of tensor quantities
on conifolds can be provided in terms of the tangent cones of
Def.(5.10). However, an explicit definition is not necessary for the
purposes of this paper as  the discussion of Einstein conifolds can be
done entirely using the techniques of completion and embedding as
utilized in the previous section.

Integration on conifolds is defined using the measure induced by the
volume element associated with the Riemannian metric $g$ at manifold
points and extending it to singular points using the distance
function.  Equivalently, this Lebesgue integral could be defined in
terms of an isometric embedding in Euclidean space. The action on a
conifold is defined in terms of Lebesgue integration of the scalar
curvature. Again, the scalar curvature is defined by extending the
scalar curvature at manifold points to the singular points. For
simplicity, the following definition assumes no boundary and
compactness:

\proclaim Definition {(6.1)}. Let $X^n$ be a closed conifold with smooth
metric $g$. The Einstein action is $$I[g] = -\frac {1}{16\pi G}
\int_{X^n} {(R-2\Lambda )}d{\mu}(g)$$ where $R$ is the scalar curvature
on $X^n -S$. \par
\noindent In the case of n-conifolds with boundary, the appropriate
boundary term is needed; as the boundary  is a (n-1)-manifold, it
follows that the required boundary term is exactly the same as that in
the corresponding n-manifold case \(hhgstate).

An {\it Einstein conifold} is a closed conifold for which the metric
$g$ restricted to $X^n-S$ is Einstein. The Euclidean Einstein equations
are elliptic in an appropriate gauge; using regularity of elliptic
partial differential equations this means the metric is analytic at
manifold points. Hence, the metric of an Einstein conifold is
particularly nice at non-singular points. The suspension of $RP^3$ with
its round metric \(idacl)  is an example of an Einstein conifold;
this example provides an excellent illustration of the analytic
properties of an Einstein conifold as discussed in section 3.
Furthermore, it will be shown that there is no loss of generality by
considering only closed conifolds when discussing Einstein conifolds by
Thm.(6.3) and Thm.(6.5).
Finally, one can show using variations with compact support on
$X^n-S$ that the following is true:

\proclaim Theorem {(6.2)}. The extremum of the action $I(g)$ on $X^n$
is an Einstein metric.

\noindent In other words, away from the singular points of the
conifold, the variation of the action yields the Einstein equations. A
detailed proof that shows that $I[g]$ is a differentiable functional
involves picking the right space of metrics and will be given
elsewhere. As conifolds become manifolds when a countable number of
points are removed, it would not be surprising if generalizations of
various useful theorems in Riemannian geometry carry over.  As
mentioned in section 3, an important result for manifolds with positive
Ricci curvature, in particular Einstein manifolds, is Myers' theorem,
Thm.(3.1).  Recall that it puts an upper bound on the diameter of the
manifold, which for a complete manifold is
is the length of a longest minimizing geodesic. Since conifolds have a
well defined notion of geodesics, the diameter can again be defined in
terms of longest minimizing geodesics. This can be used to prove a
theorem for conifolds similar to Myers' theorem:

\proclaim Theorem {(6.3)}. Let $X^n$ be a conifold with metric $g$ such
that the restriction of $g$ to $X^n - S$ is a metric with strictly
positive Ricci curvature, i.e. $R_{ab}\ge (n-1)k^2g_{ab}$ where k is a
nonzero constant. Then the diameter of $X^n$ obeys the relation
$d(X^n)\le {\pi \over k}$.

A sketch of the proof follows: First, one shows that there is no
minimizing geodesic of length greater than ${\pi \over k}$ in $X^n-S$.
To do this, assume that there is a minimizing geodesic $\gamma $ of
length $L>{\pi \over k}$.  Next, parallel transport an orthonormal
frame denoted by $\{ e_i\}_{i=1}^n$ along the curve $\gamma$ and let
$e_1$ denote the unit tangent vector of $\gamma$.  Define a new set of
$n$ vectors by multiplying the original ones by $\sin (\kappa t)$,
$w_i=(\sin (\kappa t))e_i$ where $\kappa ={\pi \over L}$ and $0\le t\le
L$ is the parameterization of $\gamma$. Since $\gamma $ is minimizing,
the first variation of the length functional of $\gamma$ must vanish,
and the second variation must be nonnegative. Explicitly calculating
the second variation of the length in terms of the vectors $w_i$ and
then summing the result from $i=2$ to $n$ implies $$\sum_{i=2}^n
D^2\ell (\gamma)(w_i,w_i)=\int_0^L {(\sin (\kappa t))}^2 [(n-1){\kappa
}^2-R_{ab}e_1^ae_1^b]dt\ . $$ Using $R_{ab}\ge (n-1)k^2g_{ab}$ and
${\kappa }^2< k^2$, it follows that $$\sum_{i=2}^n D^2\ell
(\gamma)(w_i,w_i)<0. $$ This implies that the second variation for some
$w_i$ must be negative. However, this is a contradiction because it
must be nonnegative for minimizing curves. Therefore, minimizing curves
in $X^n-S$ must have length less than or equal to ${\pi \over k}$.

The second part is to assume that there is a minimizing curve in $X^n$
with length greater than ${\pi \over k}$. If it contains no singular
points, then the argument presented above implies it cannot be a
minimizing curve. Therefore assume it contains singular points. Since
the curve is finite, it can contain at most a finite number of singular
points. One can show that by continuously perturbing the geodesic a
small amount, a new geodesic containing no singular points of length
greater than ${\pi \over k}$ can be constructed. However, no smooth
minimizing curves of length greater than ${\pi \over k}$ exist. Hence,
all minimizing curves have length less than or equal to ${\pi \over
k}$.  Therefore, $d(X^n)\le {\pi \over k}$. Q.E.D.

In particular, this theorem applies to Einstein conifolds. The bound on
the diameter is the same bound as for Einstein manifolds (see
Thm.(3.1)). Thm.(6.3) can also be combined with several simple
observations in order prove that the topology of Einstein conifolds is
restricted. Before doing so, it is necessary to define the universal
cover of a conifold.

Suppose $X$ is a path connected metric space such that every point has
a simply connected neighborhood. Then, {\it the universal covering
space} of $X$ is defined to be any space ${\tilde X}$ which is a simply
connected covering space of $X$. The universal covering space is
unique. Under the above conditions, the universal covering space can be
constructed using all continuous paths starting at some fixed point
$x_0$ in the following way: Given the set of paths
$P=\{c:[0,1]\rightarrow X|c(0)=x_0\}$, a continuous projection map
$p:P\rightarrow X$ is defined by  $p(c(t))=c(1)$. Let $\tilde X$ be $P$
modulo the equivalence relation that $c_1\sim c_2$ if and only if
$c_1(1)=c_2(1)$ and one path can be deformed continuously into the
other while holding the endpoints fixed. The projection map $p$ is also
well defined as a map on $\tilde X$, $p:{\tilde X}\rightarrow X$. Also,
the fundamental group $\pi_1(X,x_0)$ acts naturally on $\tilde X$ by
composition of paths. One can use this group action to prove that $X$
can be constructed from $\tilde X$ via identifying points under the
action of $\pi_1$. Since for any path between $x_0$ and $x\in X$, one
can add a loop in $\pi_1(X,x_0)$ to obtain a new path between $x_0$ and
$x$, it follows that $p^{-1}(x)$ has the same cardinality as
$\pi_1(X,x_0)$. Clearly, if $X$ is simply connected, then there is a
one to one correspondence between equivalence classes of paths and
points in $X$.  Hence, ${\tilde X}=X$ if and only if $X$ is simply
connected. Even if $X$ is not simply connected, by assumption each
point of $X$ has a simply connected neighborhood $U$. Using these
neighborhoods, one can show that the projection is locally a
homeomorphism, namely, each point in $X$ has a simply connected
neighborhood $U$ such that $p$ is a homeomorphism from each path
connected component of $p^{-1}(U)$ and $U$. Hence, locally $\tilde X$
looks like $X$. Finally, one can show ${\tilde X}$ as constructed is
always simply connected by verifying that ${\tilde {\tilde X}}={\tilde
X}$. Therefore, ${\tilde X}$ is the universal covering space.

Examples of universal covering spaces are  ${\bf R}^n$, which is the
universal covering space of n-torus and the group $SU(2)$, which is the
universal covering space of $SO(3)$.  Since the universal covering
space ${\tilde X}$ of a space $X$ is locally homeomorphic, the two
spaces will share many of the same local properties.  In particular, if
$X$ is a manifold, its universal cover is also a manifold.  Similarly,
as Def.(5.1) is a local definition,

\proclaim Lemma {(6.4)}. The universal covering space of a conifold is
also a conifold.

\noindent Furthermore, if the above construction is performed for a
smooth manifold and smooth curves are used, the universal covering
space will be a smooth manifold. Likewise, the same is true for smooth
conifolds.  Given any local geometric structure, it will be carried by
the universal covering space because the two spaces are locally the
same. The lemma (6.4) and this observation are crucial parts of the
next result.

\proclaim Theorem {(6.5)}. Let $X^n $ be a conifold which admits a
Riemannian metric with strictly positive Ricci curvature. Then $X^n$ is
compact.  Furthermore, $\pi_1X^n$ is a finite group.

\noindent Since the Ricci curvature is strictly positive, it follows
from the previous theorem that diameter of $X^n$ is bounded by a finite
constant.  This means that no minimizing curve can be longer than this
constant.  Recall from an earlier argument that $X^n$ can be considered
as a closed subset of Euclidean space ${\bf R}^m$ (with its geometry
induced by its embedding in Euclidean space). Hence, the diameter of
$X^n$ as a subset of Euclidean space is bounded by a constant. In other
words, $X^n$ is a closed bounded subset of Euclidean space. However,
all closed bounded subsets of ${\bf R}^m$ are compact. Therefore, $X^n$
is compact.

Let ${\tilde {X^n}}$ be the universal covering space of $X^n$. Since
the two spaces are locally equivalent, ${\tilde {X^n}}$ must admit a
metric with strictly positive Ricci curvature. Hence, ${\tilde {X^n}}$
is also compact by the above argument. Let $x\in {\tilde {X^n}}$. If
$\pi_1X^n$ is infinite, then there is an infinite sequence of distinct
points in $\tilde {X^n}$ generated by acting on $x$ with elements of
$\pi_1X^n$. Since ${\tilde {X^n}}$ is compact, the above sequence must
have a convergent subsequence $x_k$ with limit $l$. All of the points
$x_k$ in the convergent subsequence are equivalent to the same point
$x$ in the space $X^n$ because they are all constructed from each other
by acting on that point by elements of $\pi_1X^n$, thus by definition
$p(x)=p(x_k)$. Using the continuity of the projection map $p$, it
follows that $\lim_{k\rightarrow\infty}p(x_k) =p(l)$. These two
observations immediately imply that  $p(x) = p(l)$.  Hence, the limit
$l$ is equivalent to $x$ and all other $x_k$ via the group action of
$\pi_1X^n$. This means that there is an open neighborhood $U$ of $p(l)$
in $X^n$ such that $p^{-1}(U)$ is a collection of disjoint sets, one
for each member $x_k$ of the sequence. Since $l$ is the limit point of
the sequence, this is a contradiction as there can be no disjoint open
neighborhood around $l$.  Therefore, $\pi_1X^n$ must have been finite.
Q.E.D.

This theorem verifies the earlier assertion  that there was no loss of
generality by assuming that Einstein conifolds are closed. It also
obviously puts restrictions on the topology of Einstein conifolds.
Since the first homology group of a space is the abelization of its
fundamental group,  theorem (6.2) implies that the first homology is a
finite group. This means there can be no free part to the group which
implies the first Betti number is zero. Thus immediately

\proclaim Corollary {(6.6)}. Let $X^n$ be a conifold that admits a
Riemannian metric with strictly positive Ricci curvature. Then
$b_1(X^n)=0$.

\noindent This corollary is a type of Bochner vanishing theorem for
conifolds.  It yields an easy necessary condition to apply to conifolds
in order to decide if a particular conifold admits an Einstein metric.

Having developed the topology and geometry of conifolds, it is now
possible to discuss
the inclusion of  conifolds in the sum over histories formulation of
Euclidean gravity. From the earlier discussion of section 3, the
relevant question is whether or not Einstein conifolds arise in some
natural way in semiclassical quantum amplitudes.  Recall that the
motivation for the inclusion of generalized  histories lay in the
intuitive picture that there were sequences of Riemannian manifolds
with metrics that were almost Einstein that approached an Einstein
conifold; that is the Einstein
conifold arises as the limit point of a sequence of almost stationary
paths in the Euclidean gravitational integral. Indeed, given the
mathematical formulation of conifolds, this can now be proven to be the
case. The last part of this section will present the definitions used
in the proof of the main theorem and an outline of the proof. The
detailed proof of the theorem will be presented elsewhere it is rather
technically involved.

It is again useful to use the example of the suspension of $RP^3$ to
illustrate how the main result is proven. One can remove the singular
points of the  $RP^3$ conifold by removing conical neighborhoods of
each of the two singular points. The resulting space is a manifold
$I\times RP^3$ with two $RP^3$ boundaries.  As all 3-manifolds are
cobordant, the boundaries can be capped off by adding two 4-manifolds
$G$ with $RP^3$ boundaries to obtain a closed 4-manifold $M^4$. The
Einstein metric \(idacl) on the $I\times RP^3$ can be smoothly matched
to a smooth metric on each $G$ to produce a smooth metric $g$ on the
manifold. Furthermore, by removing smaller and smaller conical
neighborhoods around the singular points of the Einstein conifold and
repeating the capping off procedure, one can construct a sequence of
Riemannian manifolds $M^4_k$ each diffeomorphic to $M^4$ such the each
of the caps $G_k$ is becoming smaller with respect to the metric $g_k$.
If one chooses the sequence so that the diameter of the caps is going
to zero, then the sequence is approximating an Einstein metric on
$M^4$. Intuitively, the caps are pinching off as one takes the limit
and the diameter condition restricts the caps from pinching off other
than at a point.  The set of points on which the metrics $g_k$ on the
sequence of manifolds $M^4_k$ are non-Einstein shrinks in this limit of
zero diameter because these points are all contained within the caps.
The reason one expects to obtain the $RP^3$ Einstein conifold as the
limit of this sequence is that the caps $G_k$ are not balls for the
$RP^3$ case  but rather some complicated 4-manifold. From this
description of the convergence of the sequence of manifolds to the
$RP^3$ Einstein conifold, it is clear that the steps involved in a
general proof are to first define the sequence of manifolds that have
the needed properties and then to define what is meant by the
convergence of this sequence and finally to prove that an Einstein
conifold results from a convergent sequence of this form. Thus
\proclaim Definition {(6.7)}. A sequence of Riemannian manifolds
$(M^n_k,g_k)$ is approximately Einstein if and only if there is a
sequence of open sets $G_k$ such that \item{i)} ${\bar G}_k$ is
compact.  \item{ii)} $M^n_k=M^n$ and $G_{k+1}\subseteq G_k$.
\item{iii)} $ d_{k+1}(G_{k+1})<d_k(G_k)$.  \item{iv)} $d_k(G_k) \to 0$
as $k\to \infty$.  \item{v)} $ g_k$ is Einstein on $M^n - G_k$.\par
\noindent The following definition gives part of the needed definition
of convergence for the theorem.

\proclaim Definition {(6.8)}. A sequence of Riemannian manifolds
$\{(M^n_k,g_k)\}$ converges uniformly on compact sets to the Riemannian
manifold $(M^n_\infty,g_\infty)$ if and only if for any compact domain
$D\subseteq  M^n_\infty$
and sufficiently large $k$ there are compact domains $D_k\subseteq
M^n_\infty$ and diffeomorphisms $F_k : D\to D_k$ such that the
pullbacks $F^*g_k$ converge to the metric $g_\infty$ on $ D\subseteq
M^n_\infty$.\par

\noindent This definition can applied to the sequence of approximately
Einstein manifolds of Def.(6.7) to produce the limiting manifold with
its Einstein metric. This limiting manifold is not geometrically
complete; however the completion of this space is the desired Einstein
conifold.  It is this more general definition of convergence that is
needed in order to prove the theorem. Such a definition for any compact
metric space indeed can be provided, but it involves more technical
detail.  It can be understood if one recalls that convergent sequences
define a topology on the appropriate underlying space; it is this fact
that is used to construct the definition of general convergence for
compact metric spaces.

Given the above definitions (with the appropriate generalization) the
following theorem can be proven as described:\refto{bigmath}  \proclaim
Theorem {(6.9)}. If $\{(M^n_k,g_k)\}$ is approximately Einstein, then
$(X^n,g_\infty)$ is a conifold. Furthermore, $g_k$ converges uniformly
on compact sets to $g_\infty$ on $X^n - S$ where $S$ is the singular
set of the conifold. \par \noindent  Also note that the definition of
an approximately Einstein sequence of Riemannian manifolds and its
generalization can be directly extended to yield the definition of an
approximately Einstein sequence of conifolds. Consequently, \proclaim
Theorem {(6.10)}. If a sequence of conifolds with their metrics,
$\{(X_k^n,g_k)\}$, is approximately Einstein, then $(X^n_\infty,
g_\infty)$ is a conifold. Furthermore, $g_k$ converges uniformly on
compact sets to $g_\infty$.\par \noindent This result follows as a
trivial extension of Thm.(6.9). The full proof of these theorems will
be given in a further paper that provides the necessary results on the
convergence of the sequence.

The connection between the requirements on the sequence given in
Def.(6.7) and the result that the topology of the limit of a sequence
of manifolds is that of a conifold is made clear by observing the
relationship of Thm.(6.9) to the result of Gao on the moduli space of
Einstein metrics.\refto{gao}  Gao considers sequences of Einstein
metrics on a given fixed manifold subject to certain conditions
analogous to those of Def.(6.7) and
an additional requirement on injective radius.  Roughly speaking, the
injective radius is the size for which a normal coordinate
neighborhood is defined at each point of the manifold. The requirement
is that the injective radius of every manifold in the sequence is
bounded below by the same constant.  With such a sequence he proves
that the moduli space is compact and the topology of the manifold does
not change; the bound on the injective radius acts to keep the topology
of the manifold from pinching off in the sequence.  Thm.(6.9) has no
such restriction on the injective radius and thus the size of normal
coordinate neighborhoods can collapse and the topology of the sequence
of manifolds can be pinched off. Thus the topology of the limiting
space can differ from that of the manifolds in the sequence.  However,
the Ricci curvature and metric are still finite on the resulting
conifold.  Therefore  the  Einstein conifold is a well behaved
topological space that arises naturally as the limit of this sequence.

Further qualitative understanding of these results can be gained by
comparing them to the more familiar case of Yang-Mills theory.  The
curvature tensor is the analog of the field strength of Yang-Mills and
the moduli space of Einstein metrics like that of self dual Yang-Mills
theory is finite dimensional; thus the study of the moduli space of
Einstein metrics is analogous to the study of the space of self dual
Yang-Mills fields.  An important result in the study of Yang-Mills
theories is Uhlenbeck's theorem that singularities of self dual
Yang-Mills fields can be removed.\refto{uhlenbeck} These singularities
are purely gauge and are not coupled to the topology of the underlying
manifold. Gao's result can be considered the analog of Uhlenbeck's
theorem for Einstein gravity for a special set of manifolds with
injective radius bounded below. This special set of manifolds is needed
in order to remove topological singularities; in gravity the
singularities of the metric fields in general involve not only the
connection but the topology of the space because of the strong coupling
of topology and Riemannian geometry. Therefore a further restriction is
needed in order to make the singularities appear only in the
connection. Thm.(6.9) removes Gao's restriction; the consequence of its
removal is that topological singularities occur, in particular,
conifolds occur as boundary points of the moduli space of Einstein
manifolds. However, these boundary points still exhibit a regular
geometry.  Thus   Thm.(6.9) can be viewed as a generalization of
Uhlenbeck's theorem to Einstein gravity.

\head{{7. Conifolds as Histories in Euclidean Functional Integrals}}
\taghead{7.}

Given that Einstein conifolds occur as limit points of sequences of
approximately Einstein metrics on manifolds, one is compelled to
include their contribution to the semiclassical approximation of
Euclidean functional integrals for gravity. One expects the classical
space of histories used as a starting point for formulating Euclidean
functional integrals should be reasonably well behaved.  A space of
classical histories that does not include its limit points is not; such
spaces are not complete and generally do not exhibit the mathematical
properties expected of spaces for quantum amplitudes. Given that the
limit points of these sequences are well behaved spaces both
geometrically and topologically, it is natural to complete the space of
histories by including these points.
Thus Thm.(6.9) provides a precise mathematical motivation for
including Einstein conifolds in the semiclassical approximation.

An immediate consequence to allowing Einstein conifolds as classical
extrema of the Einstein action is that there will be semiclassical
amplitudes for a more general set of boundary topologies. An example
follows from the discussion of section 3; the Hartle-Hawking
wavefunction for $RP^3$ boundary with round metric yields a
semiclassical wavefunction. For $Ha_0<1$, there is an Einstein
conifold, the suspension of $RP^3$ with metric \(idacl) as defined in
Def.(5.5). The action of this extremum can be computed by Def.(6.1);
the curvature of \(idacl) is well defined and constant everywhere on
the suspension of $RP^3$ minus the two singular points and thus the
Lebesgue integral can be performed.  There are two possible positions
for the $RP^3$ boundary with radius $a_0$ in this solution
corresponding to filling either less than or more than half of the
conifold. Again using the prescription of  Hartle and
Hawking,\refto{hh} the Euclidean conifold extremum that dominates in
the steepest descents evaluation is that with least action
corresponding to filling less than half of the suspension of $RP^3$.
The wavefunction in the Euclidean region is thus
$$\eqalignno{\Psi_E({RP^3},a_0) &\sim \exp-{\bar I}^{-}(a_0)\cr {\bar
I}^{-}(a_0) &=  -\frac {1}{6H^2\ell^2} [(1-H^2 a_0^2)^\frac 32
-1].&(seucex)\cr}$$
where the action  is simply \(ideucex) with
$a_1=0$. It is immediately apparent from the discussion of section 3
that this wavefunction matches continuously with continuous derivative
onto $\Psi_L(RP^3,a_0)$ \(idlorex) at the point $Ha_0 =1$ for the phase
$\alpha =- \pi/4$.  Therefore this wavefunction is a semiclassical
solution of the Hartle-Hawking boundary condition when the Einstein
conifolds are allowed as Euclidean extrema.

It is clear that a similar set of semiclassical Hartle-Hawking
solutions can be constructed for any boundary of the form $S^3/\Gamma$
with round metric where $\Gamma$ is a finite subgroup of the rotation
group. However, one should note that just as not all Einstein
4-conifolds are  topologically suspensions of 3-manifolds, they are
also not geometrically such suspensions.  Secondly, although the
inclusion of  Einstein conifolds into semiclassical approximations
allows semiclassical amplitudes for a larger set of boundary
topologies, this set is still restricted.  The requirements \(kmatch)
on the intrinsic curvature of the boundary manifold  still limit its
topology  to $S^3$, $S^2\times S^1$, $S^3/\Gamma$ and connected sums of
these manifolds as discussed in section 3.  Furthermore, the geometry
of Einstein conifolds restricts the topology as well; Thm.(6.5) and its
Cor.(6.6) restrict the topology of  Einstein conifolds in a manner
completely analogous to the restriction of the topology of Einstein
manifolds by Bochner's theorem. Thus allowing Einstein conifolds as
extrema in semiclassical approximations to Euclidean functional
integrals does not radically change their properties but rather
enlarges the number of semiclassical amplitudes in a rational and
arguably desirable way.

Note that allowing Einstein conifolds as classical extrema is
self-consistent. Indeed, Thm.(6.10) proves that if one begins with a
sequence of approximately Einstein conifolds, one does not end up with
a more general topological space, but rather another Einstein
conifold.  This self-consistency obviously does not occur for sequences
of approximately Einstein manifolds by Thm.(6.9). Therefore the
inclusion of Einstein conifolds leads to a complete moduli space and is
thus a reasonable extension to the moduli space of Einstein manifolds.

Finally, given the inclusion of Einstein conifolds in the space of
classical Euclidean solutions,  it is eminently reasonable to propose
the set of compact conifolds as generalized histories for Euclidean
functional integrals for gravity.  Clearly, topological spaces that
occur as extrema  of the Euclidean action are suitable as spaces for
more general smooth histories as well. In addition, the appropriate set
of such topological spaces is the set of all conifolds rather than just
those that admit Einstein metrics; the arguments given in section 2 for
the need to include all manifolds  as histories can be extended to show
that a similar set of conifolds must be used as well.  For example, the
generalized  Hartle-Hawking wavefunction \(hhgstate) is expressed as
$$\eqalignno{\Psi[\Sigma^{n-1}, h] &= \sum_{X^n}\int Dg
\exp\biggl(-I[g] \biggr)\cr I[g] &=- \frac 1{16\pi G} \int_{X^n}
(R-2\Lambda) d\mu(g)
 - \frac 1{8\pi G}\int_{\Sigma^{n-1}}  K d\mu(h) &(congstate)\cr}$$
where the mathematical description  of  the histories is: A {\it
generalized history} is a pair $(X^n,g)$ where $X^n$ is a smooth
compact conifold and $g$ is at least a $C^2$ metric on $X^n-S$ with the
specified induced metric $h$ on the boundary $\Sigma^{n-1}$. This
definition includes all Riemannian histories. Like Riemannian
histories, these generalized histories are classical histories of the
theory and consequently, one expects that they provide the underlying
topology for an appropriate set of distributional histories for the
Hartle-Hawking wavefunctional.

One consequence of allowing conifold histories in the Euclidean
functional integral is that the principle of equivalence, that is that
spacetime is locally ${\bf R}^n$,
is no longer automatically enforced by the set of histories in the
sum. Therefore, as emphasized by Hartle,\refto{unruly} it is now
possible to consider the issue of whether or not the principle of
equivalence holds for a given quantum amplitude calculated from the
generalized sum over histories.  As the principle of equivalence is a
property of classical Lorentzian spacetime, it is reasonable to expect
it to appear only in quantum amplitudes corresponding to classical
spacetimes.  In order to fully address this issue, one must have a
method of determining when a given quantum amplitude corresponds to a
classical spacetime. Therefore this issue is closely tied to that of
interpreting quantum amplitudes for the universe. This question of the
interpretation of the quantum mechanics of gravity is complex and
unresolved and this paper will not touch on it.\refto{interpret}
However, given the close connection of classical solutions and
semiclassical wavefunctions, it is useful to provide a  discussion of
principle of equivalence in the context of
semiclassical Hartle-Hawking wavefunctions.

There are two simplistic methods of associating a classical spacetime
with a given semiclassical amplitude. The first is to associate the
stationary path used to construct the semiclassical amplitude with a
classical spacetime. The second is to use the the semiclassical
wavefunction itself to provide initial data for the classical
spacetime. In both approaches, classical spacetime is only associated
with a complex extremum of the Euclidean action; such complex extrema
are typically Lorentzian solutions to the Einstein equations. Therefore
questions about the principle of equivalence can only be asked in
regions of configuration space $(\Sigma^{n-1},h)$ where the
semiclassical wavefunction is formed from such extrema.

In the first approach, the question about the principle of equivalence
can be rephrased into two questions, are there Lorentzian spacetimes
which exhibit non-manifold singularities and do these spacetimes occur
in the semiclassical approximation to the Euclidean functional integral
for a given boundary geometry. The answer to the first question is well
known to be yes; it follows immediately from the singularity theorems.
In fact, explicit examples of singular Lorentzian spacetimes with
isolated singular points can be constructed.  The answer to the second
question is maybe; in order to answer this question, one has to decide
whether or not a Lorentzian solution to the Einstein equations with
such a nonmanifold singularity is an complex extrema to the Euclidean
functional integral \(congstate). To do so, one must provide a set of
regularity conditions on the curvature and metric suitable for
application in Lorentzian conifolds similar to those given for
Euclidean conifolds. Given such a set of regularity conditions, one
could then see if
there was or was not a singular Lorentzian solution for the specified
boundary geometry. In any case, the important point to stress is that
if such a Lorentzian spacetime exists, it also exists classically.
Therefore, the properties of the spacetime derived from these
semiclassical amplitudes are completely determined by the classical
theory. Therefore, the question about whether or not the principle of
equivalence holds is really a question about a particular classical
Lorentzian spacetime.

In the second approach, regions of classical spacetime correspond to
regions of configuration space for which the quantum amplitude becomes
oscillatory.  Classical spacetime is retrieved from the semiclassical
limit of the wavefunction $\Psi[\Sigma^{n-1},h] \sim \cos(S[h] +
\alpha)$ as the evolution along the normals to the surfaces of constant
phase $ \pi_{ij} = \frac {\delta S}{ \delta h_{ij}}.$  That is the
metric and its conjugate momenta
 $(\pi, h)$ derived from a given wavefunction provide the initial data
for a family of Lorentzian spacetimes with topology $\Sigma^{n-1}\times
R$. Given sufficiently regular initial data, it is well known that it
can be evolved for a finite distance.  In the semiclassical
approximation, $\pi$ is a continuous differentiable tensor field for
actions evaluated on continuous differentiable solutions of the
Einstein equations. Thus the principle of equivalence holds for
semiclassical wavefunctions  for which the complex stationary path is
sufficiently regular.  Note that this means that the principle of
equivalence holds automatically for any semiclassical wavefunction
where the initial data is appropriate initial data for a Lorentzian
spacetime, even if there are nonmanifold points to the past of the
boundary $(\Sigma^{n-1},h)$ either in the Lorentzian or Euclidean
regions.  Of course, the evolution of this initial data may result in a
Lorentzian solution containing nonmanifold points to the future.
Whether or not it does depends on the explicit form of the
 initial data. Therefore, whether or not the principle of equivalence
holds in a classical spacetime associated with a given semiclassical
wavefunction is again equivalent to the same question for Lorentzian
solutions of the Einstein equations.

Thus, the issue of whether or not the principle of equivalence holds in
the classical limit for a set of generalized histories is completely
determined by the properties of the Lorentzian solutions themselves in
these simplistic interpretations. Note that both of these simplistic
approaches completely ignore the issue of how probable a given
Lorentzian spacetime is; in fact it follows from the above discussion
that it is this issue that is really the one of interest.  It is clear
that both a more detailed computation of the quantum amplitude and a
more sophisticated interpretation of the amplitude is needed to really
address this issue. In any case, Euclidean quantum amplitudes using
generalized histories provide a viable starting point for such a study
because in the most simplistic interpretation, their semiclassical
limit corresponds to Lorentzian spacetimes.

\head{8.~Conclusion}

This paper proposes a new set of topological spaces called conifolds
and argues  that the set of smooth compact
 conifolds form a suitable and necessary set of topological spaces for
generalized histories for Euclidean functional integrals for gravity.
It can be proven that Einstein conifolds arise as the limit of a
sequence of approximately Einstein manifolds.  Thus Einstein conifolds
correspond to the boundaries of the  moduli space of Einstein
manifolds. Additionally, sequences of approximately Einstein conifolds
also converge to Einstein conifolds.  Therefore it is natural to
include such spaces as histories for Euclidean functional integrals for
gravity.

The immediate benefit of using conifold histories  in Euclidean
functional integrals for gravity such as \(congstate)
  is that semiclassical amplitudes corresponding to Einstein conifolds
follow immediately. However there are additional benefits as discussed
in Part II of this paper.  As the set of conifolds is larger than the
set of manifolds,
  Euclidean sums over histories for gravity formulated using conifolds
have the additional advantage of being algorithmically decidable in
four or fewer dimensions. Thus, unlike for the case of sums over
manifold histories, these sums can be implemented in a systematic way.
Additionally, the sums can be explicitly carried out in finite
approximations to expressions such as \(congstate) formulated in terms
of Regge calculus in four dimensions. Obviously many difficulties with
the formulation of Euclidean integrals for Einstein gravity will
 remain for any generalization of the histories to any set of more
general topological spaces that includes all manifolds. However, these
problems are not any more severe for the set of conifolds than for
manifolds. Moreover, the topological issues addressed are relevant to
many theories involving sums over topological spaces. Therefore the
above proposal provides a starting point for addressing further issues
regarding the formulation of Euclidean integrals for gravity.

\subhead{\undertext{Acknowledgments}}

This work is the outgrowth of results presented by the authors at the
Fifth Marcel Grossman meeting in Perth, Australia, 1988 and elsewhere.
The authors would like to thank the relativity group at the University
of Maryland for their hospitality during the time this work was
initiated.  This work was supported in part by the Natural Science and
Engineering Research Council of Canada, the NSERC International
Fellowship Program and CITA.

\head{Appendix A}
\subhead{\undertext{Mayer-Vietoris}}

Given a topological space built from the union of other topological
spaces, a useful result for calculating the relation of the homology of
the pieces to that of the whole space is the
Mayer-Vietoris theorem.

\proclaim {Theorem (A.1)}. Given a topological space $Y$ and two open
subsets $U$ and $V$ such that $Y=U\cup V$ then the following sequence
of homology groups is exact

$$\eqalignno{\dots\dar H_k(U\cap V)\aar H_k(U) &\oplus H_k(V)\btar
H_k(Y) \dar H_{k-1}(U\cap V)\aar \dots \cr \dar H_1(U\cap V)\aar
H_{1}(U)\oplus H_{1}(V)\btar H_{1}( &Y)\dar H_0(U\cap V)\aar
H_0(U)\oplus H_0(V)\btar H_0(Y)\cr}  $$
where the homology groups are taken to have
coefficients in any abelian group.

Recall that exact means that the composite of any two homomorphisms in
the above sequence is zero, e.g. $\alpha_*\beta_* =0$, and the kernel
each homomorphism is equal to the image of the previous one, e.g.
ker($\beta_*$) = im($\alpha_*$).
If real coefficients are used, the homology groups are vector spaces
over the real numbers and the above exact sequence is a sequence of
vector spaces. If the integer coefficients are used, the homology
groups are direct products of vector spaces over the integers and
cyclic groups.

The Mayer-Vietoris sequence can be applied to show that $RP^3$ cannot
be embedded in $S^4$ or $RP^4$ so that it divides them in half.  The
homology groups of all of these manifolds can be computed (for example
via the application of the Mayer-Vietoris sequence to a decomposition
of the manifold); thus two of the three homology groups in the sequence
are known. This information is sufficient to determine the homology
groups of the remaining space. Namely,

\proclaim {Theorem (A.2)}. There is no compact topological 4-manifold
$M$ with boundary $RP^3$ so that $M\cup M$ is $S^4$ or $RP^4$.

First, consider the case of $RP^4$. For this case it is sufficient to
use real homology.  Assume a manifold $M$ as above exists. Writing the
Mayer-Vietoris sequence starting at $H_4(RP^4)$ gives $$H_4(RP^4)\ar
H_3(RP^3)\aar H_3(M)\oplus H_3(M)\btar H_3(RP^4) \ . $$ Since $RP^4$ is
nonorientiable, $H_4(RP^4)=0$ and by  an explicit calculation,
$H_3(RP^4)=0$ with real coefficients.  Also, $H_3(RP^3)={\bf R}$
because $RP^3$ is orientiable. Hence, the exact sequence is $$0\ar {\bf
R}\aar H_3(M)\oplus H_3(M)\ar 0 \ . $$ This is a contradiction because
the vector space ${\bf R}$ is one dimensional and it cannot be
decomposed into the sum of two identical vector spaces.  Therefore,
$RP^3$ can not divide $RP^4$.

Second, consider the case of $S^4$. A similar contradiction can be
obtained by using the lower part of the Mayer-Vietoris sequence with
integer coefficients $$H_2(S^4)\dar  H_1(RP^3) \aar H_1(M)\oplus
H_1(M)\btar H_1(S^4) \ . $$ Now $H_1$ of any space is the abelization
of its fundamental group. Since $\pi_1(RP^3)={\bf Z}_2$ and $S^4$ is
simply connected, $H_1(RP^3)=Z^2$ and $H_1(S^4)=0$. Additionally, by
explicit calculation, $H_2(S^4)=0$.
Hence, $$0\dar {\bf Z}_2 \aar H_1(M)\oplus H_1(M)\btar 0 \ . $$ This
is a contradiction because a cyclic group of order two can not be the
direct sum of two groups.  Therefore, $RP^3$ does not divide $S^4$.
Q.E.D.

The Mayer-Vietoris sequence can be used to prove a useful relationship
between the Euler characteristic of a manifold and that of its boundary
in odd dimensions.  \proclaim {Theorem (A.3)}. Let $M$ be an compact
manifold of odd dimension $n$ and with boundary $\partial M$. Then
$$2\chi(M) = \chi(\partial M).$$\par \noindent First, form the manifold
$N = M\cup M$ by doubling over $M$ on its boundary $\partial M$. By
construction, $N$ is a closed connected odd dimensional manifold.
Applying Thm.(A.1) to this decomposition of $N$ using real coefficients
yields $$\eqalignno{0\dar H_n(\partial M)\aar 2H_n(M)\btar H_n(N) \dar
&H_{n-1}(\partial M)\aar  2H_{n-1}(M)\btar H_{n-1}(N)  \dots \cr \dar
H_1(\partial M)\aar 2H_{1}(M)\btar H_{1}( &N)\dar H_0(\partial M)\aar
2H_0(M) \btar H_0(N)\cr}  $$ where the factors of $2$ in the above
denote the sum of the two identical vector spaces, $2H_q(M) =H_q(M)
\oplus  H_q(M)$.
 It can be proven that the alternating sum of the dimensions of the
vector spaces $H_k$ in the Mayer-Vietoris sequence sum to zero;
consequently $$\sum_{k=0}^n (-1)^k[\hbox{\rm dim}(H_k(\partial M)) -
2\hbox{\rm dim}(H_k(M) )+ \hbox{\rm dim}(H_k(N))] = 0.\eqno(dim)$$ Now
$H_k(\partial M) = 0$ for $k>n-1$ as  $n-1$ is the dimension of the
boundary $\partial M$. Also the betti numbers $b_k=\hbox{\rm dim}H_k$
and the alternating sum of the betti numbers is the Euler
characteristic. Thus \(dim) is equal to $ \chi(\partial M) - 2\chi(M) +
\chi(N) = 0$. Finally, note that the Euler characteristic of closed odd
dimensional manifolds is zero. Thus $\chi(\partial M) = 2\chi(M)$.
Q.E.D.

\references

\refis{rs} C.~P.~Rourke and B.~J.~Sanderson,
{\it Introduction to Piecewise-Linear Topology},\hfill\break
(Springer-Verlag, New York, 1972).

\refis{dw} D.~M.~Witt, {\sl Phys. Rev. Lett.,} {\bf 57}, 1386, (1986).

\refis{donandjim} J.~B.~Hartle and D.~M.~Witt, {\sl Phys. Rev. D,}
 {\bf 37}, 2833, (1988).

\refis{h1} S.~W.~Hawking, in {\it General Relativity: An Einstein
Centenary Survey} edited by S. W. Hawking and W. Israel
(Cambridge University Press, Cambridge, 1979).

\refis{hh} J.~B.~Hartle and S.~W.~Hawking, {\sl Phys. Rev. D,}
{\bf 28}, 2960, (1983).

\refis{h2} S.~W.~Hawking, {\sl Nucl. Phys. B,} {\bf 239}, 257, (1984).

\refis{diffgeom} See for example M.~W.~Hirsch, {\it Differential Topology},
(Springer-Verlag, New York, 1976).

\refis{freedman} M.~Freedman and F.~Quinn, {\it Topology of 4-Manifolds},
 (Princeton University Press, Princeton, 1990).

\refis{ghp} G.~W.~Gibbons, S.~W.~Hawking and M.~J.~Perry,
{\sl Nucl. Phys.B,}
{\bf 138}, 141 (1978), K.~Schleich, {\sl Phys. Rev. D,}
{\bf 36}, 2342, (1987).

\refis{jimhall} See for example, J.~J.~Halliwell and J.~B.~Hartle,
{\sl Phys. Rev. D,} {\bf 41}, 1815, (1990), J.~J.~Halliwell and J.~Louko,
 {\sl Phys. Rev. D,} {\bf 42}, 3997, (1990), and references
theirin.

\refis{interpret} The issue of interpretation is old, controversial and
unresolved. See for example M.~Gell-Mann and J.~B.~Hartle in
{\it Complexity, Entropy and the Physics of Information}, ed. W.~Zurek
(Addison-Wesley, Reading, 1990), D.~Page, {\sl Phys. Rev. D,} {\bf 34},
2267, (1986), and B.~S.~DeWitt,
{\sl Phys. Rev.,} {\bf 160}, 1113, (1967).

\refis{continuity} Continuity of the wavefunction and its derivative
also follows directly from the presumed equivalence of the sum over
histories formulation and the Wheeler de Witt equation. The Wheeler de
Witt equation  $ {\bf H} \Psi(h) = 0$ is a functional equation where
${\bf H}= \square + { }^3R(h) - 2\Lambda$ is the operator corresponding
to the Hamiltonian constraint of the theory, \(lconstraints). The
kinetic term $\square$ is formally the wave operator with respect to
the metric on superspace; thus the Wheeler de Witt equation is formally
a hyperbolic second order differential equation.  In minisuperspace
models, this equation is finite dimensional. Given a finite dimensional
second order hyperbolic differential equation of the form of the
Wheeler de Witt equation and differentiable initial data, one can
easily show that a solution exists in a neighborhood of the initial
hypersurface and is also differentiable.  In particular, semiclassical
solutions corresponding to Euclidean instantons provide such
differentiable initial data by the analyticity properties of Einstein
manifolds. Therefore one can extend the semiclassical wavefunction from
areas of configuration space where there is a Euclidean stationary
point to areas where there is not. Thus it follows that the
wavefunction and its derivative must be continuous for minisuperspace
models, at least in a neighborhood of the Euclidean region. It is also
reasonable to expect the same in general, at least for semiclassical
solutions; a proof for the functional equation of course requires a
definition of the functional derivative, a concrete definition of the
Wheeler de Witt operator, and an appropriate choice of function space,
all of which are problematic.

\refis{toy} R.~Kirby and L.~Siebenmann, {\it Foundational Essays on
Topological Manifolds, Smoothings, and Triangulations},
(Princeton University
Press, Princeton, 1977).

\refis{kn} A good standard reference for the  differential
geometry of manifolds is S.~Kobayashi and K.~Nomizu,
{\it Foundations of Differential Geometry}, vols. 1 and 2,
(Wiley, New York, 1963).

\refis{jb} See, for example W.~A.~Wright and I.~G.~Moss,
{\sl Phys. Lett. B,}
{\bf 154}, 115, (1985), P.~Amsterdamski, {\sl Phys. Rev. D,}
{\bf 31}, 3073, (1985), J.~J.~Halliwell and S.~W.~Hawking
{\sl Phys. Rev. D,} {\bf 31}, 1777, (1985),
S.~W.~Hawking and D.~N.~Page, {\sl Nucl. Phys. B}
{\bf 264}, 185, (1986).

\refis{anderson} M.~T.~Anderson, {\sl J. Amer. Math. Soc.}
{\bf 2}, 456, (1989).

\refis{gao} L.~Z.~Gao, {\sl J. Diff. Geom.} {\bf 32}, 155, (1990).

\refis{II} K.~Schleich and D.~M.~Witt, ``Generalized Sums
over Histories II:
Simplicial Conifolds'', UBC preprint, 1992.

\refis{bigmath} K.~Schleich and D.~M.~Witt, in preparation.

\refis{IIa} For a complete list of references to the
literature on algorithmic
decidability and classifiability, please see Reference [8].

\refis{spanier} E.~H.~Spanier, {\it Algebraic Topology},
(McGraw-Hill, New York, 1966).

\refis{nash} J.~Nash, {\sl Ann. of Math. (2)} {\bf 63}, 20, (1956).

\refis{uhlenbeck} K.~Uhlenbeck, {\sl Comm. Math. Phys.},
{\bf 83}, 11 (1982).

\refis{glimm} See for example M.~Reed and B.~Simon, {\it Methods of
Modern Mathematical Physics II: Fourier Analysis, Self-Adjointness},
(Academic Press, New York, 1975), J. Glimm and A.~Jaffe,
{\it Quantum Physics:
A Functional integral Point of View}, (Springer-Verlag, New York, 1981).

\refis{footnote1} If the space has a triangulation, this sum can
 be expressed as the alternating
sum of vertices, edges, triangles, and tetrahedra in
the triangulation.

\refis{thurston} W.~Thurston, {\sl Bull. Amer. Math. Soc},
{\bf 6}, 357, (1982).

\refis{genlist} See for example, S.~Giddings and A.~Strominger,
 {\sl Nucl.~Phys. B,} {\bf 306},  890, (1988) and  S.~Coleman,
{\sl Nucl.~ Phys. B,} {\bf 310}, 643, (1988), J.~Polchinski,
{\sl Nucl. Phys. B,} {\bf 325}, 619, (1989).

\refis{besse} A.~ Besse, {\it Einstein Manifolds},
(Springer-Verlag, New York, 1987).

\refis{geroch} R.~Geroch, {\sl J.~Math.~Phys.} {\bf 8}, 782, (1967).

\refis{unruly} J.~B.~Hartle, {\sl Class.~Quant.~Gravity} {\bf 2}, 707,
(1985).

\endreferences
\vfill\eject

\head{Figure Captions}

\noindent {\bf Figure 1:} The product space $X\times Y \times I$ and
the join of $X$ and $Y$ obtained by identifying all points in  $Y$ at
$X \times Y \times \{0\}$ and all points in $X$ at $X \times Y \times
\{1\}$.

\bigskip

\noindent {\bf Figure 2:} The suspension of $RP^2$ represented by a
solid 3-ball with identifications. Each cross section of the ball is a
2-ball with the indicated identifications on its $S^1$ boundary.

\bigskip

\noindent {\bf Figure 3:} A sequence of slicings of the suspension of
$RP^3$. The  $RP^3$ manifold is represented by a solid 3-ball subject
to the condition that antipodal points on  its 2-sphere boundary are
identified. Note that no identification occurs on interior points of
the 3-ball.

\bigskip

\noindent {\bf Figure 4:}
Two illustrations of polyhedra with badly behaved curvatures: two
2-spheres connected along a line segment and two cones with their
apexes identified.

\bigskip

\noindent {\bf Figure 5:}
A 4-conifold produced by removing three disks from a 2-manifold of
genus 4, taking the product of the result with a 2-sphere and then
coning off the boundaries. Two dimensions have been suppressed; each
nonsingular point is a 2-sphere  but the singular points $a$, $b$, and
$c$ are just points.

\vfill\eject\vfill\supereject\end